\documentclass{jpp}
\usepackage{graphicx}

\usepackage[utf8]{inputenc}
\usepackage[T1]{fontenc}
\usepackage{amsmath}
\usepackage{amssymb}
\usepackage{dsfont}

\newcommand{\sgn}{\operatorname{sgn}}

\usepackage{unicode}
\newcommand{\csso}{\fontencoding{LECO}\selectfont\char215}
\newcommand{\iotaslash}{\iota\hspace*{-0.45em}\text{\csso}}

\renewcommand\Im{\mbox{$\mathrm{Im}$}}

\newcommand{\vpar}{\ensuremath{v_{\parallel}}}
\newcommand{\vperp}{\ensuremath{v_{\perp}}}
\newcommand{\xpar}{\ensuremath{x_{\parallel}}}
\newcommand{\xperp}{\ensuremath{x_{\perp}}}

\providecommand\bnabla{\boldsymbol{\nabla}}
\providecommand\bkappa{\boldsymbol{\kappa}}

\providecommand\bnabla{\boldsymbol{\nabla}}

\def \pa {\mbox{$\partial$}}
\def \al {\mbox{$\alpha$}}

\def \kpar {\mbox{$k_{\parallel}$}}
\def \kperp {\mbox{$k_{\perp}$}}
\def \Lpar {\mbox{$L_{\parallel}$}}
\def \vth {\mbox{$v_{\mathrm{T}}$}}
\def \vp {\mbox{$\varphi$}}
\def \vt {\mbox{$\vartheta$}}

\def \bl {\mbox{$b(\ell)$}}
\def \blp {\mbox{$b(\ell^{'})$}}
\def \bhat {\mbox{$\hat{b}$}}

\def \blp {\mbox{$b(\ell^{'})$}}
\def \l {\mbox{$\ell$}}
\def \lp {\mbox{$\ell^{'}$}}

\def \eps {\mbox{$\epsilon$}}
\def \lampar {\mbox{$\lambda_{||}$}}

\renewcommand \o {\mbox{$\omega$}}

\def \os {\mbox{$\omega_*$}}
\def \osT {\mbox{$\omega_*^{\mathrm{T}}$}}

\def \ost {\mbox{$\widetilde{\omega}_*$}}

\def \op {\mbox{$\omega_{\parallel}$}}
\def \od {\mbox{$\omega_d$}}

\def \odt {\mbox{$\widetilde{\omega}_d$}}

\def \odb {\mbox{$\overline{\omega}_{d}$}}
\def \ob {\mbox{$\overline{\omega}$}}
\def \odtav {\mbox{$\left<\odt \right>$}}
\def \odav {\mbox{$\left<\od \right>$}}

\def \Reff {\mbox{$R_{\mathrm{eff}}$}}

\def \th {\mbox{$\theta$}}

\def \gzh {\mbox{$\hat{\Gamma_{0}}$}}
\def \goh {\mbox{$\hat{\Gamma_{1}}$}}

\shorttitle{ITG linear critical gradient}
\shortauthor{G. T. Roberg-Clark,  G. G. Plunk and P. Xanthopoulos}

\title{Calculating the linear critical gradient for the ion-temperature-gradient mode in magnetically confined plasmas}

\author{G. T. Roberg-Clark\aff{1}
  \corresp{\email{gar@ipp.mpg.de}},
  G. G. Plunk\aff{1}
 \and P. Xanthopoulos\aff{1}}

\affiliation{\aff{1}Max-Planck-Institut F\"ur Plasmaphysik, D-17491, Greifswald, Germany}

\begin{document}

\maketitle

\begin{abstract}
 A first-principles method to calculate the critical temperature gradient for the onset of the ion-temperature-gradient mode (ITG) in linear gyrokinetics is presented. We find that conventional notions of the connection length previously invoked in tokamak research should be revised and replaced by a generalized correlation length to explain this onset in stellarators. Simple numerical experiments and gyrokinetic theory show that localized "spikes" in shear, a hallmark of stellarator geometry, are generally insufficient to constrain the parallel correlation length of the mode. ITG modes that localize within bad drift curvature wells that have a critical gradient set by peak drift curvature are also observed. A case study of nearly helical stellarators of increasing field period demonstrates that the critical gradient can indeed be controlled by manipulating magnetic geometry, but underscores the need for a general framework to evaluate the critical gradient. We conclude that average curvature and global shear set the correlation length of resonant ITG modes near the absolute critical gradient, the physics of which is included through direct solution of the gyrokinetic equation. Our method, which handles general geometry and is more efficient than conventional gyrokinetic solvers, could be applied to future studies of stellarator ITG turbulence optimization.
\end{abstract}

\section{Introduction} \label{sec:intro}

Energy transport resulting from small-scale electrostatic fluctuations is a major impediment to sustaining nuclear fusion in magnetic confinement devices. While one can scale up the size of a reactor to offset these losses and reach the required fusion product, smaller designs with better transport properties are more likely to be built and to succeed against economic competition from other energy sources. Thus precise control of the level of electrostatic fluctuations would be a boon to the fusion program. The problem of transport resulting from electrostatic turbulence has been tackled aggressively over the past decades in the context of axisymmetric tokamaks and is receiving increasingly more attention in the stellarator community. The complex magnetic geometry of stellarators makes the problem more difficult to study, but also opens the door to possible optimization by exploring the large theoretical space of stellarator designs. The current hope is that once the magnetic field shape is adjusted to confine trapped particle orbits, there will be enough remaining freedom to optimize for turbulence; see e.g. \cite{Mynick2006a} for a review.

The Ion-Temperature-Gradient-Drive-Mode (ITG mode or ITGM) has been singled out as a leading cause of transport in magnetic fusion devices. This mode, which is driven by gradients in ion temperature, causes heat losses that act to reduce the steep gradient imposed by heating in the core of the device. Several different approaches to modeling the ITGM in stellarators have emerged over the past decade, such as incorporating the effect of zonal flows into quasilinear turbulence saturation \citep{Nunami2013,Toda2019}, adding the contributions of subdominant eigenmodes into quasilinear saturation estimates \citep{Pueschel2016a}, and calculating mode saturation by energy transfer to damped modes \citep{Terry2006,Hatch2011}. Some optimization strategies based on nonlinear modeling have also been identified e.g. targeting key geometric quantities associated with ITG intensity \citep{Mynick2010a,Xanthopoulos2014a} and increasing the correlation time of unstable modes with damped modes \citep{Hegna2018}. Renewed interest in near-axis expansions for stellarators \citep{Landreman2018,Jorge2020a} has also led to calculation of the geometric features that influence the ITG mode \citep{Jorge2020b,Jorge2021a} with possible application to optimization for ITG turbulence. Much of the work just mentioned, as well as current efforts towards turbulence optimization that we are aware of, involve modeling turbulence itself. This is a challenging problem to crack, especially at the level of generality needed for stellarator design. We can solve a simpler problem that applies to any toroidal configuration, however, by finding the ITG linear critical gradient.

The critical gradient is the threshold gradient for onset of the ITG mode. While some plasma instabilities are susceptible to sub-critical turbulence that bring this onset below the linear threshold, it appears that the opposite occurs in the case of the ITGM, with the with onset occurring at a somewhat larger value of the gradient \citep{Dimits2000}. The critical gradient is also a valuable reference point since the thermal transport above this value tends to be ``stiff'' in tokamaks, i.e. it often sharply increases with the gradient. If the plasma has a radial temperature profile with a constant gradient that matches the critical gradient, $d\ln{T}/dr=d\ln{T}/dr_{crit}=\text{const}$ ($T$ is the temperature and $r$ is the radial distance from the core of the plasma), one can infer from integrating this profile inward in $r$ that the temperature increases exponentially towards the core. It stands to reason that a modest improvement in the critical gradient can result in a significant gain in core temperature. A specific gradient might be targeted so that modest turbulence is present in the desired operating regime, which may, in certain cases, help flush out impurities \citep{Garcia-Regana2021a}.

\subsection{Physics of the ITG mode critical gradient}

The critical gradient has a long history in tokamak research. Early analytical works mapped the stability of the ITGM in parameter space and calculated threshold gradients in various limits [e.g. \cite{Kadomtsev1995a,Terry1982,Romanelli1989a,Biglari1989a,Hahm1989a,Dominguez1988,Dominguez1989a,Romanelli1990}]. For a comprehensive study of the analytical calculations and their validity the reader is referred to \cite{Zocco2018}, in which the critical gradient is explored in the more general cases of low-shear tokamaks and stellarators in the so-called local kinetic limit. The critical gradient has also been studied for the electron-temperature-gradient (ETG) mode [e.g. \cite{Horton1988a,Jenko2001a,Jenko2002a}], whose behavior mirrors the ITG in certain limits. Although the ITG critical gradient for the actual onset of turbulence will differ from the linear value, it is only expected to be greater due to an ``upshift'' owing to zonal flows. That is to say, the true critical gradient in many devices is, to first approximation, set by linear physics and is improved to some degree by nonlinear effects. 

The physics of the ITG is simpler near the linear threshold. The complexity introduced by multiple growing modes having the same perpendicular wavenumber is removed, as only one marginally unstable mode remains, the ``last mode standing''. More importantly, the prediction of the linear critical gradient does not require a complete understanding of turbulence, and is universally relevant in the sense that virtually all configurations of interest are expected to have a well-defined value. The critical gradient is thus a convenient metric for comparing how susceptible different configurations are to ITG turbulence. However, there is no general method for calculating the critical gradient in arbitrary geometry aside from running a series of linear gyrokinetic simulations essentially as a root finder, which is a tedious and computationally intensive procedure. This motivates the development of an efficient and robust method, which is the main result of this paper. We also try to clarify some outstanding physics questions concerning what controls the stability of the ITG mode in general magnetic geometry.

We distinguish here between two linear critical gradients, corresponding to the onset of the two distinct branches of the ITG mode. For configurations with low global shear, it is the threshold of slab-like modes, spread broadly along the field line, that characterizes the absolute critical gradient, below which no unstable modes are present. Such ``background'' modes and the dependence of their critical gradient on the temperature ratio of ions and electrons were studied by \cite{Zocco2018}, with further evidence for slab-like modes in low-shear stellarators discussed in works such as \cite{Faber2018}. Sufficiently large global shear is expected to stabilize the slab-like background and lead to a single critical gradient, the onset of resonant modes related to the toroidal branch of the ITG mode whose critical gradient is set by drift curvature. These curvature-influenced modes (that still retain the parallel resonance associated with the slab mode) ``balloon'' and localize along the field line, where they, at least in part, feed off of regions of bad curvature along the field line. The onset of these modes defines a second critical gradient, which is thought to be set by the strength of bad curvature, as well as the extent of regions of bad curvature along the field line. Both slab-like background and localized toroidal/slab modes are expected to emerge in low to moderate shear devices, leading to a ``knee'' in the plot of maximal growth rate versus temperature gradient (see \cite{Zocco2018}). Which of these critical gradients is more important in setting the onset of significant heat transport remains an open question.

The traditional geometric scale invoked to understand the onset and drive of the ITG mode has been the connection length, typically the distance along the field line between the inboard and outboard sides of a tokamak (the distance between regions of ``good'' and ``bad'' curvature). If an ITG mode is in the toroidal branch, it will typically balloon in the center of the bad curvature well. Near the transition to the slab mode (if one exists as in the low-shear case), it may be that the mode has to spread out to fill the entire bad curvature region, i.e. connection length, in order to be unstable. In the case of a stellarator, one might attempt to generalize the connection length idea, for example, to be the distance between sharp features in local shear that are inherent to non-axisymmetric fields that also tends to correspond to the scale of variation of the normal curvature. We will show that this is not completely successful in describing the critical gradient. Fortunately, a more fundamental measure is available, namely the ion {\em correlation} length, or the distance over which ion motion is correlated with the mode. Analysis of the equations will show that this is simply a properly defined characteristic frequency divided by the ion thermal velocity. Thus we find that for stellarators the traditional notion of connection length (defined for curvature by local shear, etc.) should generally be replaced by this correlation length. The ITG mode near marginality may, depending on the size of this length scale, be driven independently of local features in the geometry and, in such cases, depends on average properties along a magnetic field line.

The paper is structured as follows. In section \ref{sec:definitions} we define the linear gyrokinetic equation and the integral form of the equation that we use to solve for the ITG critical gradient. We discuss the local, linear theory of the slab ITG mode critical gradient in section \ref{sec:kadomtsev} and extend the discussion to the effects of field-line-varying geometry in section \ref{sec:gyrotunnel}. Numerical experiments using a slab geometry model show that it is difficult to constrain the correlation length of the ITG with localized amplification of the perpendicular wavenumber in section \ref{sec:shearspikes}. These experiments illustrate an example of possible conflict between the concepts of correlation length and connection length. We also provide evidence for the onset of curvature-driven modes by extending the numerical model to include a spatially-varying curvature in section \ref{sec:curv-onset}. A case study for increasing the critical gradient in nearly-helically symmetric toroidal configurations (section \ref{sec:helical}) shows that attempts to reduce the curvature connection length with large toroidal field period numbers leads to a critical gradient instead set by global shear. We are thus motivated to confront the first-principles calculation of the critical gradient in section \ref{sec:critgrad-solve}, the main result of the paper. Section \ref{sec:conclusion} concludes the paper.

\section{Linear electrostatic gyrokinetic equation}\label{sec:definitions}

\subsection{Definitions}

Following \cite{Plunk2014a}, we let $g_{i}(\vpar,\vperp,\mathbf{x},t)$ be the non-adiabatic part of $\delta f_{i}$ and $\phi(\mathbf{x})$ be the electrostatic potential, with $\mathbf{x}$ the gyro-center position. Here the $i$ subscript refers to a single ion population and $\delta f_{i}$ is the gyro-averaged perturbation of the distribution function from the equilibrium Maxwellian distribution (defined below). The independent velocity coordinates are parallel $(\vpar)$ and perpendicular ($\vperp$) to the equilibrium magnetic field $\mathbf{B}(\mathbf{x})$. We use the twisted slicing representation \citep{Roberts1965}, $(g_{i},\phi)=(\hat{g}_{i}(l),\hat{\phi}(l))\exp(i s-i \omega t)$, where $\l$ is the magnetic field-line-following coordinate (arc length variable) defined such that $\bhat = \mathbf{B}/|\mathbf{B}|=\partial \mathbf{x}/\partial \l$. $\o$ is the Fourier-expanded frequency, $t$ is time, and $s(\mathbf{x})$ is the eikonal factor containing variation in the spatial directions perpendicular to $\mathbf{B}$. The factor $\exp(i s)$ contains fast spatial oscillations with the anisotropy condition $\mathbf{B} \cdot \bnabla s=0$ ensured by taking $\partial s/\partial l = 0$. Representing the magnetic field in flux coordinates, $\mathbf{B}=\bnabla \psi \times \bnabla \al$, we set 

\begin{equation}\label{eqn:kperp}
    \bnabla s \equiv \mathbf{k_{\perp}} = k_{\alpha} \bnabla \alpha + k_{\psi} \bnabla \psi,
\end{equation}
where $k_{\alpha}$ and $k_{\psi}$ are constants and the variation of $\mathbf{k_{\perp}}(l)$ comes from the geometric quantities $\bnabla \alpha$ and $\bnabla \psi$. Here $\psi$ is a flux surface label and $\alpha$ is a label for a particular field line on a given surface $\psi$. Thus the gyro-center position is expressed as $\mathbf{x}=\mathbf{x}(\psi,\alpha,\l)$. Defining a toroidal angle $\zeta_{\text{tor}}$ and poloidal angle $\theta_{\text{pol}}$ we also write $\alpha=\theta_{\text{pol}}-\iotaslash \zeta_{\text{tor}}$, with $\iotaslash=\iotaslash(\psi)$ the rotational transform. These definitions allow us to re-express eq. (\ref{eqn:kperp}) as

\begin{equation}\label{eqn:kperp-shear}
    \mathbf{k}_{\perp}=k_{\alpha}(\bnabla \theta_{\text{pol}} - \iotaslash \bnabla \zeta_{\text{tor}} - \frac{d \iotaslash}{d \psi}\zeta_{\text{tor}} \bnabla \psi)+k_{\psi}\bnabla \psi. 
\end{equation}

We associate the term proportional to $d \iotaslash / d\psi$ with global shear, which leads to a secular increase of $|\mathbf{k}_{\perp}|$ as $\zeta_{\text{tor}}$ or $\ell$ are increased. The remaining, non-secular terms proportional to $k_{\alpha}$ are then considered to be local shear effects.

From now on we drop the $i$ subscript and hat notations, retaining electron ``e'' subscripts for two subsequent definitions. We then write the linear gyrokinetic equation for the ions as
\begin{equation}
i\vpar \frac{\partial g}{\partial \ell} + (\o - \odt)g = \varphi J_0(\o - \ost)f_0\label{gk-eqn}
\end{equation}\label{eqn:lingyro}

\noindent with the following definitions: $J_0 = J_0(k_{\perp}v_{\perp}/\Omega) = J_0(k_{\perp}\rho\sqrt{2}v_{\perp}/\vth)$; the thermal velocity is $\vth = \sqrt{2T/m}$ and the thermal ion Larmor radius is $\rho = \vth/(\Omega\sqrt{2})$; $n$ and $T$ are the background ion density and temperature; $q$ is the ion charge; $\varphi = q\phi/T$ is the normalized electrostatic potential; $\Omega=q B/m$ is the cyclotron frequency, with $B=|\mathbf{B}|$ the magnetic field strength.  Assuming Boltzmann electrons, the quasineutrality condition is

\begin{equation}
\int d^3{\bf v} J_0 g = n(1 + \tau) \varphi,\label{qn-eqn}
\end{equation}\label{eqn:poisson}

\noindent where $\tau = T/(ZT_e)$ with the charge ratio defined as $Z = q/q_e$. The equilibrium distribution is the Maxwellian

\begin{equation}
f_0 = \frac{n}{(\vth^2\pi)^{3/2}}\exp(-v^2/\vth^2),
\end{equation}

\noindent and we introduce the velocity-dependent diamagnetic frequency

\begin{equation}
\ost = \osT \left[\frac{v^2}{\vth^2} - \frac{3}{2}\right]
\end{equation}

\noindent where we neglect background density variation and define $\osT = (Tk_{\alpha}/q)d\ln T/d\psi$.  The magnetic drift frequency is $\odt = {\bf v}_d\cdot{\bf k}_{\perp}$ and the magnetic drift velocity is ${\bf v}_d = \hat{\bf b}\times((v_{\perp}^2/2)\bnabla \ln B  + \vpar^2\bkappa)/\Omega$, where $\bkappa = \hat{\bf b}\cdot\bnabla\hat{\bf b}$. We take $\bnabla \ln B = \bkappa$ (small $\beta$ approximation) for simplicity. We then let

\begin{equation}
\odt = \frac{\mathbf{k_{\perp}} \cdot (\bhat \times \bkappa)v^{2}_{T}}{\Omega} \left[\frac{\vpar^2}{\vth^2} + \frac{\vperp^2}{2\vth^2}\right] = \od(\l) \left[\frac{\vpar^2}{\vth^2} + \frac{\vperp^2}{2\vth^2}\right],
\end{equation}

\noindent where the velocity-independent drift frequency $\od(\ell)$ generally varies along the field line. 

\subsection{Integral form of the equation}

An integral equation can be derived from Eqns.~\ref{gk-eqn}-\ref{qn-eqn} assuming ``outgoing'' boundary conditions $g(\vpar > 0, \ell = -\infty) = g(\vpar < 0, \ell = \infty) = 0$, consistent with ballooning modes that decay as $|\ell| \rightarrow \infty$ \citep{Connor1980,Romanelli1989a}.  To enforce these conditions we assume the system has non-zero global shear, $d\iotaslash/d\psi \neq 0$, though it is allowed to be small. As shown in appendix \ref{ballooning-appx}, one then obtains

\begin{align}
(1+ \tau)\varphi(\ell) = \frac{-2i}{\vth\sqrt{\pi}}\int_{0}^{\infty} \frac{d\xpar}{\xpar} \int_{0}^{\infty}  d\xperp \xperp (\o - \ost) J_0 \nonumber \\ 
\times \int_{-\infty}^{\infty} d\ell^{\prime} J_0^{\prime} \exp(-x^2 + i \sgn(\ell-\ell^{\prime})M(\ell^{\prime}, \ell))\varphi(\ell^{\prime}),\label{eqn:ballooning-disp}
\end{align}

\noindent where $\xperp = \vperp/\vth$ and $\xpar = \vpar/\vth$, $\sgn$ gives the sign of its argument, $J_0 = J_0 \left(\sqrt{2b(\ell)}\xperp \right)$, $J_0^\prime = J_0 \left(\sqrt{2b(\ell^\prime)}\xperp \right)$, and $b(\ell)=\rho^{2}k_{\perp}^{2}(\ell)$. The physics of the drift resonance is contained in the factor

\begin{equation}\label{eqn:mbar}
M(\ell^{\prime}, \ell) = \int_{\ell^{\prime}}^{\ell} \frac{\o - \odt(\ell^{\prime\prime})}{\vth \xpar } d\ell^{\prime\prime}.
\end{equation}

We neglect particle trapping so $\xpar$ and $\xperp$ do not depend on $\ell$. Most evidence indicates that the ITG mode uniformly responds to changes in the parameters $\tau$ and $\mathrm{d}n/\mathrm{d}\psi$ and is stabilized by increasing either of them, except for a relatively small region of parameter space where positive density gradients can destabilize the mode. For simplicity we thus set $\tau=1$ and $\bnabla n=0$. 

\section{Physics of marginally unstable ITG modes} \label{sec:GENEmodels}

\subsection{Slab ITG mode}\label{sec:kadomtsev}

The historical emphasis on connection length can be motivated by the local kinetic slab ITG critical gradient from \cite{Kadomtsev1995a}. In the local limit the equilibrium quantities do not vary along $\l$. The Kadomtsev \& Pogutse result can be derived by taking the linear gyrokinetic equation (\ref{eqn:lingyro}), assuming $\partial / \partial l \rightarrow i \kpar$, and combining it with Poisson's equation (\ref{eqn:poisson}) to yield the linear dispersion relation. The physics of marginal stability is here connected to the parallel slab (or Landau) resonance condition $\o - \kpar \vpar=0$, finite Larmor radius effects entering the linear dispersion relation through the $\kperp$ dependence of the $J_{0}$ factors in eqs. (\ref{eqn:lingyro}) and (\ref{eqn:poisson}), and the drive term $\osT$ which also depends on $k_\alpha$. As reviewed in \citet[Eq. 10]{Plunk2014a} if one then takes the limit $\gamma \rightarrow 0+$, where $\gamma$ is the imaginary part of the mode frequency $\omega$, a threshold gradient for destabilization of the mode can be derived,

\begin{equation}\label{eqn:kadomtsev}
\frac{\osT}{\os} = \eta = \frac{1}{1+2b(1-\Gamma_{1}/\Gamma_{0})} \left[1+\sqrt{1+\frac{2(1+\tau)(1+\tau-\Gamma_{0})}{\Gamma^{2}_{0}\os^{2}/\omega^{2}_{||}}(1+2b(1-\Gamma_{1}/\Gamma_{0}))} \right],
\end{equation}
where the density gradient has been kept such that $\eta=\mathrm{d} \ln{T}/\mathrm{d}\ln{n}$, $\Gamma_{0,1}(b)=\exp(-b)I_{0,1}(b)$, $I_{0,1}(b)$ is the modified Bessel function of order $0$ or $1$, $b=k^{2}_{\perp}\rho^{2}$, and $\op=k_{||}v_{T}$. Finite $k_{\psi}$ tends to stabilize ITG modes as it does not enter the drive term $\osT$ and can amplify the effect of shear in $\kperp$. In some cases finite $k_{\psi}$ can be destabilizing for the ITG mode if this reduces the FLR stabilization by $k_{\alpha}$ along the flux tube. For simplicity we set $k_{\psi}=0$. We then write

\begin{equation}\label{eqn:osTradial}
\osT=k_{\alpha}\frac{T}{q} \frac{d \ln T}{dr} \frac{dr}{d\psi}\equiv \frac{k_{\text{pol}}\rho}{2} \left(\frac{v_{T}}{L_{T}}\right),
\end{equation}
with the radial coordinate $r$ defined through $\psi=(1/2)B_{0}r^2$, $B_{0}$ is a constant such that $\rho=mv_{T}/(qB_{0})$, and in the last step the poloidal wave number $k_{\text{pol}}=B_{0}k_{\alpha}(dr/d\psi)=k_{\alpha}/r$ and temperature gradient length scale $L_{T}\equiv (d \ln T/dr)^{-1}$ have been inserted. We let $k^{2}_{\perp}\rho^2=k^{2}_{\text{pol}}\rho^2=b$, again for simplicity, and take the limit $\os / \op \rightarrow 0$, solving for $\osT / \op$ (note $\osT = \eta \os$). Letting $\kpar \rightarrow 2\pi / \lampar$, where $\lampar$ is the parallel wavelength of the mode, we arrive at
\begin{equation}\label{eqn:gabekadomtsev}
        \frac{\lampar}{L_{T}}=4\pi \sqrt{\frac{2(2-\Gamma_{0})}{b \Gamma_{0}(\Gamma_{0}+2b(\Gamma_{0}-\Gamma_{1}))}}=4\pi F(b),
\end{equation}
with $\tau$ set equal to $1$. The absolute critical temperature gradient is found by minimizing $F(b)$, resulting in $b_{\text{crit}}\simeq 0.88$ and $(4\pi)^{-1}\lampar/L_{T\text{crit}}\simeq 2.6$. When $b \gtrsim 1$, $F$ is increased through finite Larmor radius effects (i.e. the modified Bessel function factors damp the mode) while for $b<b_{\text{crit}}$ the drive from $\osT \simeq \sqrt{b}$ is weakened, also increasing $F$ and hence the linear critical gradient.

One can interpret $\lampar$ as not just the mode extent but also the parallel correlation length $\Lpar$, the distance a thermal ion travels in the characteristic timescale $1/\o \sim 1/\osT$. One could then estimate the damping rate associated with decorrelation as $v_{T}/\Lpar$. The simple scaling $1/L_{T\text{crit}} \propto 1/\Lpar$ can also be inferred from eq. (\ref{eqn:gabekadomtsev}), implying that shortening the correlation length could be an effective strategy for increasing the ITG linear critical gradient.

\subsection{Tunneling of the ITG Mode}\label{sec:gyrotunnel}

It is logical to conclude that if the correlation length $\Lpar$ can be imposed by magnetic field geometry, through e.g. local amplification of $\kperp$ or the size of a curvature well, then one could increase the ITG critical gradient to a desired level by controlling the magnetic field shape. Before testing this idea in slab geometry (section \ref{sec:shearspikes}) we look at the integral equation (\ref{eqn:ballooning-disp}) to make qualitative arguments about the effect of shear. These arguments and subsequent numerical tests reveal that equating $\Lpar$ and the traditional connection length is not generally correct.

\cite{Waltz1993} have argued that local shear and curvature in stellarators may play dominant roles in setting the radial extent and growth rate of the ITG mode. Their analysis was carried out in the cold ion fluid limit. Although local shear is thus expected to have a stabilizing effect, we find that modes are not completely confined by large, local amplification factors of $\kperp$ in the gyrokinetic limit.

For marginally stable modes with $\gamma = \Im [\omega] \rightarrow 0+$ the velocity integrals in (\ref{eqn:ballooning-disp}) can be strongly suppressive (i.e. causing the integrand of the $\ell^\prime$ integral to be small) at large enough $\lp-\l$. Mode suppression by shear occurs when $\kperp(\ell$ or $\ell^{'})$ becomes large and different from one other because of the field-line variation of $k_{\perp}$. We expect this to occur generally when the so-called shear amplification factor $f=\kperp^{2}(\l)/\kperp^{2}(\l=0)$ reaches values much larger than $1$. This leads to a mismatch in the arguments of the two oscillatory Bessel function factors $J_{0}J_{0}^{'}=J_{0}(\sqrt{2\kperp (\l)}\xperp)J_{0}(\sqrt{2\kperp(\lp)}\xperp)$. When the $\xperp$ integration is performed the overall integral (\ref{eqn:ballooning-disp}) is then suppressed. In analogy with quantum mechanics we picture the ITG mode as a wavefunction that is attempting to penetrate a steep potential barrier when it encounters a region of large $k_\perp(\l)$. We think of segments of the field line with large $\kperp$ as forbidden regions, through which the eigenfunction $\varphi(\l)$ would need to tunnel in order to access the temperature gradient drive from further regions along the field line. 

Another ingredient in the tunneling picture is the ion distribution function, which accumulates phase-space structure as it propagates through the large $k_\perp$ barriers. The accumulation can be inferred from the equation \ref{g-soln-1}, in which the Bessel function factors $J_{0}J^{'}_{0}$ and phase factor $M$ will rapidly contribute phase space structure in a manner similar to phase mixing. However, like in phase mixing, this process is technically reversible and can be partially undone if $k_{\perp}(\l)$ is of a similar magnitude on both sides of the barrier (which is exactly what occurs with local shear). That is, $\varphi(\l)$ may be transiently suppressed in a narrow region and then recover outside this region, due the smoothing or "unwinding" of the phase-space structure of $g$. As a result, large, fluctuating shear amplification factors (which one may associate with local shear "spikes" in stellarators) may not be as effective in localizing the extent and drive of the ITG mode as previously thought. If $\kperp(\l)$ has a secularly growing component, however, phase space structure can continue to accumulate in $g$. In this case the the amplitude of $\varphi(\l)$ must inevitably decay at large $\l$, along with the drive of the ITG mode. Global shear (see equation \ref{eqn:kperp-shear}) therefore plays a unique role in the stability of the ITG mode because it causes a secular increase of $\kperp$ along the field line.

We mention here a less relevant case in which modes have $k_{\perp}\lesssim 1$ on the entire field line (which can occur for zero global shear) and for which the Bessel functions cannot suppress the mode amplitude. If this is the case then the mode is never confined by shear and is essentially a free slab mode that violates ballooning boundary conditions. This is why we require nonzero global shear to capture ballooning modes in equation (\ref{eqn:ballooning-disp}). In the next section we study a sheared slab model to explore the local confinement and tunnelling phenomena in more detail.

\subsection{Numerical experiments using GENE}\label{sec:shearspikes}

We take the work in \cite{Plunk2014a} as a starting point for our numerical experiments. In that paper the authors solved the ITG linear dispersion relation of "boxed" modes in a flux tube geometry with uniform geometric coefficients except at the boundaries, where they set $k_{\perp}=\infty$ at locations $\l=-L_{0}/2$ and $\l=L_{0}/2$, with $L_{0}$ the width of a box in $\l$. We study a similar problem but the modes here are instead confined by $\kperp$ barriers of finite width and height, where $k_\perp(\l)$ has no secularly growing component, i.e. we only retain non-global shear terms (see equation \ref{eqn:kperp-shear}). Linear flux-tube simulations are performed with the GENE gyrokinetic code \citep{Jenko2000a} to find the ITG critical gradient in a sheared slab (i.e. with curvature set to $0$). Defining the coordinates $(x,y,z)$ similarly to \cite{Faber2018} we set $x=(\psi-\psi_{c})/(B_{0}r_{c})$, $y=r_{c}(\alpha-\alpha_{c})$ and $z=\l$. Here $\psi_{c}$ and $\alpha_{c}$ are the toroidal flux and field line labels, respectively, at the center of the flux-tube simulation domain with small spatial extents perpendicular to the magnetic field line. These extents are defined as $-\Delta \psi \leq \psi - \psi_{c} \leq \Delta \psi $ and $-\Delta \alpha \leq \alpha - \alpha_{c} \leq \Delta \alpha $. $r_{c}=\sqrt{2\psi_{c}/B_{0}}$ is the radius corresponding to the center of the simulation domain. The perpendicular wavenumbers (with no field-line dependence included yet) are $k_{x}= m\pi B_{0} r_{c}/\Delta \psi$ and $k_{y}=u \pi/(r_{c}\Delta \alpha)$, with $m$ and $u$ integers. The field-line variation of $k_{\perp}$ is expressed through $k^{2}_{\perp}(\l)=k^{2}_{x}g^{xx}(\l)+k_{x}k_{y}g^{xy}(\l)+k^{2}_{y}g^{yy}(\l)$, with the metric coefficients defined as $g^{xx}(\l)=(\bnabla x)^2(\l)$, $g^{xy}(\l)=(\bnabla x)(\l)\cdot (\bnabla y)(\l)$, and $g^{yy}(\l)=(\bnabla y)^2(\l)$. Finally, the perpendicular wavenumbers and coordinates are normalized with respect to $\rho$, such that $\kperp \rho$ and the metric coefficients are dimensionless. 

\begin{figure}
  \centering
  \includegraphics[scale=0.45]{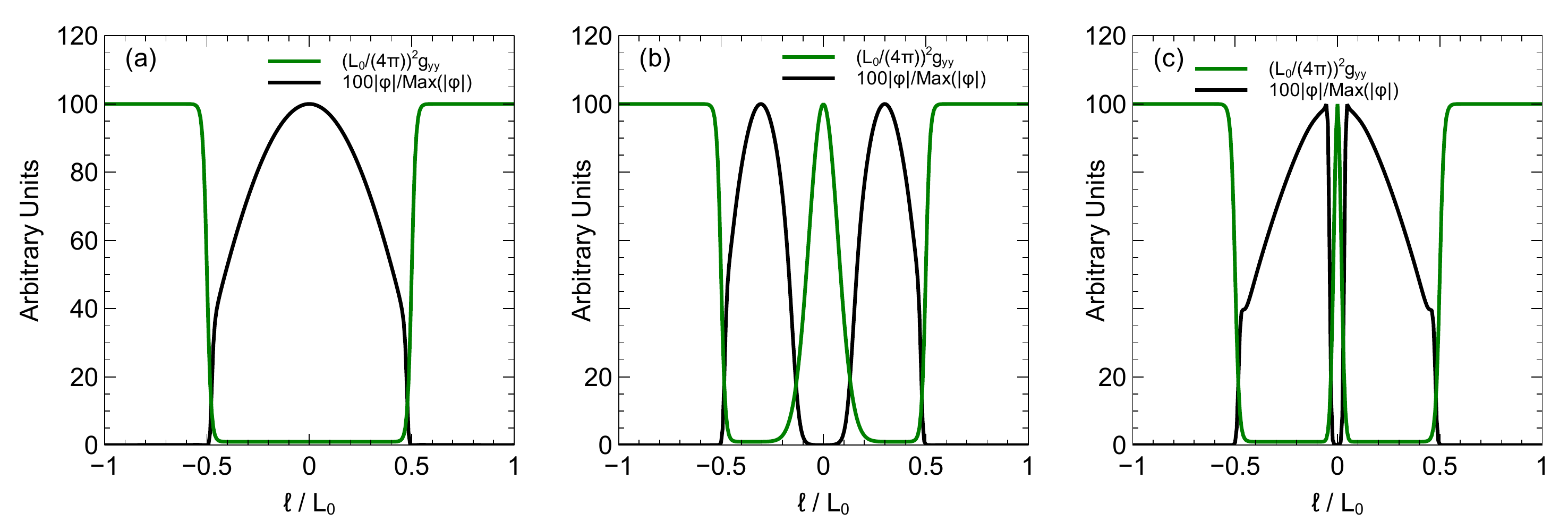}
  \caption{$\ell$ profiles of $g_{yy}$ (green curves) and $|\vp|$ (black curves) of the marginally unstable slab ITG mode ($\gamma \rightarrow 0+$) for simulations with (a) only bounding walls, (b) walls and a wide middle spike, and (c) walls with a narrow middle spike. See Table \ref{tab:shearspikes} for simulation parameters and critical gradients. $|\vp|$ is normalized to its maximum value and multiplied by 100 to match $g_{yy}$.}
\label{fig:shearspikes}
\end{figure}

By setting the amplification factor $f=k^{2}_{\perp}(\l)/k^2_{\perp}(\l=0)$ we mimic the effect of magnetic shear when simulating the growth of the ITG mode, in this case through a localized "spike" and bounding walls in the profile of $k_\perp$. Assuming $m=0$ for simplicity such that $k_{x}=0$ and $k^{2}_\perp=k^{2}_{y}g^{yy}$, we set $g^{yy}=(\bnabla y)^{2}$ by hand,
\begin{equation}\label{eqn:shearspikes}
    g^{yy}(\ell)=1+
    99\left(1+\frac{1}{2}\left(\text{Tanh}\left[\frac{16\pi}{L_{0}}\left(\ell-\frac{L_{0}}{2} \right) \right]
    -\text{Tanh}\left[\frac{16\pi}{L_{0}}\left(\ell-\frac{L_{0}}{2} \right) \right] \right)  \right) + H e^{-\ell^{2}/w^{2}}.
\end{equation}
The width of the box between the bounding walls is $L_{0}$ while the localized Gaussian spike has variable amplitude $H$ and width $w$. In the $H=0$ case the profile is that of a constant $\kperp$ middle region that smoothly but rapidly transitions to a constant $f=g^{yy}(\ell)/g^{yy}(0)=100$. When $H\neq0$ the Gaussian spike causes a large change in the derivative of $\kperp$ near $\l=0$, which quickly reverses sign, causing $\kperp$ to return to its value in the flat region on the other side of the spike. This type of rapid but transient increase in $\kperp$ is what we mean when we refer to local shear. Electrons are adiabatic, $T_{e}=T$, there is no density gradient, and $|\mathbf{B}|$ is constant. The parallel boundary condition is set by using twist-and-shift \citep{Beer1995a} with zero connections such that the electrostatic potential $\vp$ is $0$ at the boundaries. $\vp$ is well-behaved near the boundary (has no hints of mode activity) provided the walls are sufficiently wide and high. We also exclude modes with such low values of $k_{y} \rho$ that they cannot be confined by barriers with shear amplification factors that lack a secularly growing component (as mentioned in section \ref{sec:gyrotunnel}) and present results for $k_{y}\rho >0.4$, which are shown in Fig. \ref{fig:shearspikes} with corresponding data in Table \ref{tab:shearspikes}. The critical gradient is found by running a series of simulations, starting with large $L_{0}/L_{T}$, and steadily reducing the drive until only one unstable $k_{y}\rho$ mode remains which satisfies $\varphi(\l)=0$ at the boundaries, determining both the critical $(k_{y}\rho)_\text{crit}$ and $L_{0}/L_{T\text{crit}}$.

\begin{table}
  \begin{center}
\def~{\hphantom{0}}
  \begin{tabular}{lccccc}
      Fig. \ref{fig:shearspikes} label & $H$  & $w/L_{0}$ & Area $=H w/((99)(0.1) \, L_{0})$ &  $(4\pi)^{-1}L_{0}/L_{T\text{crit}}$ & $(k_{y }\rho)_\text{crit}$\\
     (a) &  0    & 0       & 0 & $2.67$ & $0.8$\\
     (b) & 99    & $0.1$   & 1 & $6.28$ & $0.5$\\
         & 99    & $0.07$  & 0.7 & $4.71$ & $0.7$\\
         & 99    & $0.05$  & 0.5 & $4.39$ & $0.7$\\
     (c) & 99    & $0.025$ & 0.25 & $4.08$ & $0.8$\\
         & 56.25 & $0.1$   & 0.75 & $5.49$ & $0.5$\\
         & 25    & $0.1$   & 0.5  & $5.03$ & $0.6$\\
         & 6.25  & $0.1$   & 0.25 & $3.77$ & $0.5$\\
          
  \end{tabular}
  \caption{Slab ITG simulation results}
  \label{tab:shearspikes}
  \end{center}
\end{table}

In figure \ref{fig:shearspikes}(a) we show $g^{yy}(\ell)$ and $|\vp(\ell)|$ in the simplest case of only bounding walls, $H=0$, for a simulation with $(k_{y}\rho)_\text{crit} =0.8$ near marginality, $\gamma=4\pi \times 10^{-3} c_{s}/L_{0}$. The critical gradient is $(4\pi)^{-1}L_{0}/L_{T\text{crit}}\simeq 2.67$ in good agreement with the Kadomtsev-Pogutse formula (eq. (\ref{eqn:gabekadomtsev}), $(4\pi)^{-1}L_{0}/L_{T\text{crit}}\simeq 2.60$), though the simulation result yields a smaller $b_{\text{crit,sim}} = (k_{y}\rho)^2_\text{crit} = 0.64$ than that of the formula ($b_{\text{crit}}\simeq 0.88$). The difference in $b_{\text{crit}}$ is not surprising in light of the fact that the $\l$-periodic slab solution from the Kadomtsev-Pogutse result is qualitatively different from mode subject to ballooning boundary conditions. The parallel connection length inferred by the critical gradient still agrees decently well between the two cases.

We now vary $H$ and $w$. For $w/L_{0}=0.1$, Fig. \ref{fig:shearspikes}(b) shows the division of $L_{0}$ into two regions for the new marginal ITG mode, whose critical gradient has more than doubled (see Table \ref{tab:shearspikes}) and whose amplitude $|\vp(\ell)|$ goes to zero within the central spike. One can therefore still use (\ref{eqn:gabekadomtsev}) to estimate the critical gradient with $\lampar=L_{0}/2$, the distance between the spike and a wall. As we shall soon find, however, this estimate becomes unreliable for the case of weaker shear features.

If the spike height is maintained while its width is narrowed to $w/L_{0}=0.025$ (fig. \ref{fig:shearspikes}(c)), tunneling becomes significant and the mode can maintain strong correlations across the spike. Here $\vp(\ell)$ still approaches $0$ for $|\ell|/w \lesssim 1$ as a result of the large factor $f$ as in Fig. \ref{fig:shearspikes}(b). The mode amplitude quickly recovers outside the spike and for $|\ell|/w \gtrsim 1$ closely resembles that of the $H=0$ case. The critical gradient is about $1.5$ times larger than that of the $H=0$ case in contrast with the factor $>2$ increase seen in the $w/L_{0}=0.1$ case. This means the parallel correlation length of the mode is now closer to the original $L_{0}$ and the wall-to-spike distance no longer matches this length. To effectively reduce the ITG mode correlation length, therefore, a localized amplification $f$ must not only be large in amplitude but also have significant width, i.e. it must have area along the field line. We can thus interpret the spike as reducing the correlation length by carving out larger or smaller pieces of the domain in Figs. \ref{fig:shearspikes}(b) or (c) respectively, yielding larger or smaller increases in the critical gradient relative to the reference case with $H=0$ (Fig. \ref{fig:shearspikes}(a)). Results of scans in which $H$ and $w$ were gradually reduced (Table \ref{tab:shearspikes}) suggest a roughly linear scaling of the critical gradient with spike area.

The required shear amplification factors $f\sim 100$ used in these test cases are fairly large compared to those of typical low-global-shear stellarators. We refer here to the work of \cite{Jorge2020b}, who plot the field-line variation of quantities important to ITG mode stability such as $(\bnabla \alpha)^{2}$ and components of drift curvature $\od$ for many standard stellarator configurations (Figures 4-13 of that paper). Inspection by eye suggests $(\bnabla \alpha)^2/(\bnabla \alpha^2)|_{\l=0}$ for these flux tubes is in the range of $5$ to $40$. 

We caution that our model does not capture all the features of a consistent equilibrium, in particular, the modulation of the drift frequency from local shear that can affect the stability of the ITG mode. The results in this section nonetheless suggest that it is difficult to significantly increase the critical gradient with localized spikes in $\kperp$.

\subsection{Curvature-driven ITG modes}\label{sec:curv-onset}

\begin{figure}
  \centering
  \includegraphics[scale=0.45]{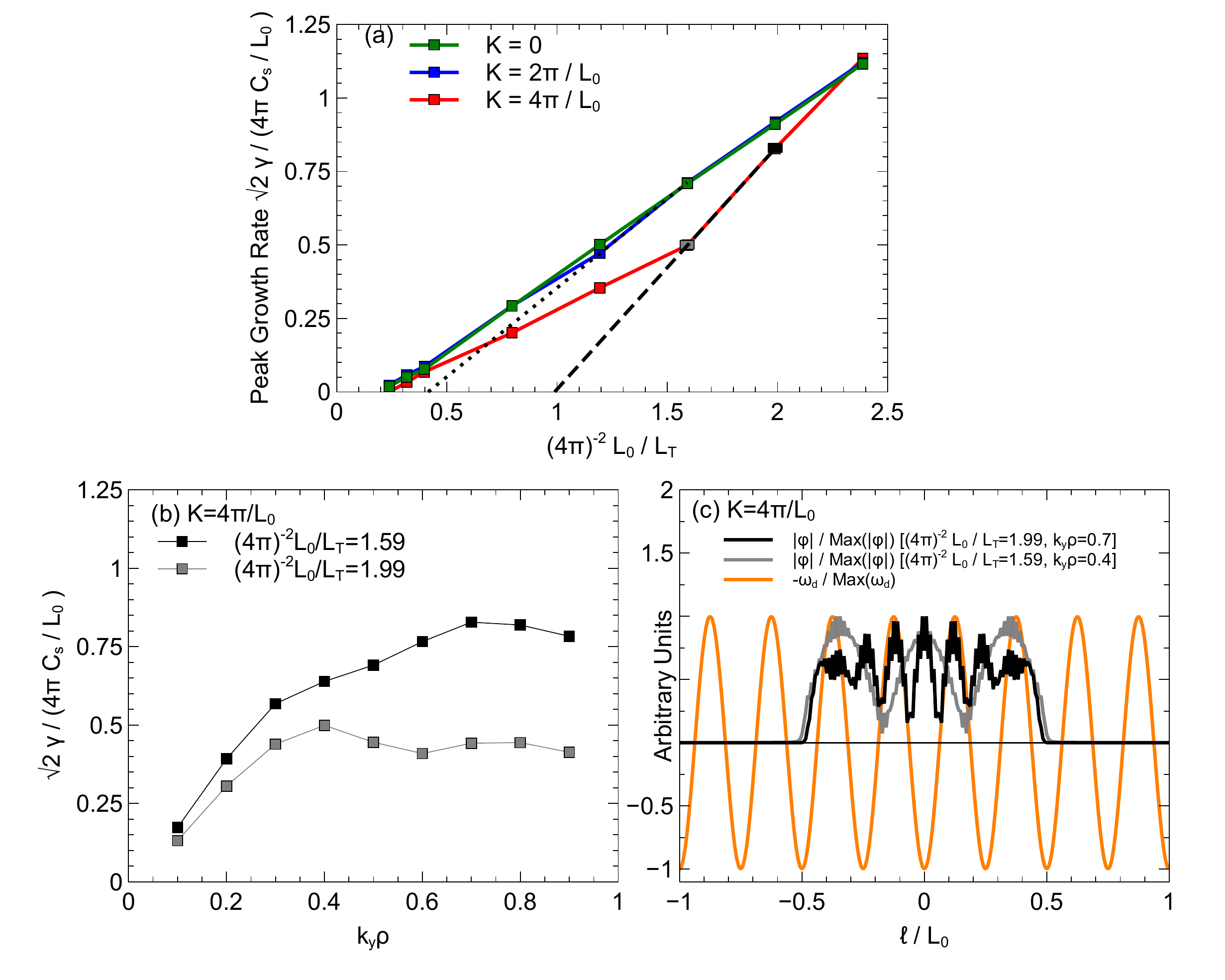}
  \caption{Onset of curvature-driven resonant ITG modes in a simple geometry model using GENE. The $g^{yy}$ profile in (\ref{eqn:shearspikes}) is used with a drift curvature profile $\od/(k_{y}\rho)=K\cos[8\pi \l/L_{0}]$ added. (a) Peak growth rate $\gamma$ versus $L_{0}/L_{T}$ for $K=0$ (green curve), $K=2\pi/L_{0}$ (blue curve, weaker curvature), and $K=4\pi/L_{0}$ (red curve, stronger curvature). A strong ``knee'' is visible in the growth rate near $(4\pi)^{-2}L_{0}/L_{T}=1.59$ for the red curve. Data points are shown with squares. Linear extrapolation to an inferred critical gradient set by curvature is shown with a dashed black line. The two data points used in this extrapolation correspond to the points before (silver square, $(4\pi)^{-2}L_{0}/L_{T}=1.59$, peak growth at $k_{y}\rho=0.4$) and after (black square, $(4\pi)^{-2}L_{0}/L_{T}=1.99$, peak growth at $k_{y}\rho=0.7$) the transition in mode structure and growth rate of the most unstable modes as $L_{0}/L_{T}$ is increased. A weaker knee is visible near $(4\pi)^{-2}L_{0}/L_{T}=1.2$  on the horizontal axis for the blue curve, with a dotted black line for its inferred critical gradient. (b) Growth rate spectra $\gamma(k_{y}\rho)$ before and after the mode transition for $K=4\pi/L_{0}$ discussed above. (c) Mode profiles $|\varphi(\ell)|$ for the two cases presented in (b) at peak growth rate, $k_{y}\rho=0.4$ (silver curve) and $k_{y}\rho=0.7$ (black curve). The orange curve shows the normalized drift frequency profile for the cases with curvature.}
\label{fig:curv-onset}
\end{figure}

ITG modes can also be excited by the drift resonance, $\osT \sim \od$, rather than the parallel resonance $\osT \sim v_{T}/\Lpar$ discussed in section  \ref{sec:kadomtsev}. Although the local ITG linear dispersion relation including $\od$ must be solved numerically [see e.g. \cite{Plunk2014a} and references therein], it is well known from analytic theory considerations that damping of the mode can occur when the effect of curvature is too large [e.g. \cite{Sugama1999a}]. That is, when $\od \gtrsim \osT$, the mode is pulled out of the drift resonance and decorrelation occurs. The marginal stability criterion becomes $1/L_{T\text{crit}} \propto 1/R_{\text{eff}}$, where $\Reff=\kperp \rho v_{T}/(\sqrt{2}\od)$ is the effective radius of curvature and implies a parallel correlation length $\Lpar \sim v_{T}/\o \sim v_{T}/\od$ for curvature-driven ITG modes. This shows why larger values of bad curvature will actually increase the critical gradient. Once the mode ``balloons'' and localizes to the curvature well, however, it will be driven more strongly beyond the threshold if curvature is increased. Therefore we expect that, when peak bad curvature is increased, the growth rate should become more stiff (have a larger slope versus gradient) above the second critical gradient. We confirm these intuitions in the following numerical experiments.

We again take the shear profile (\ref{eqn:shearspikes}) of just bounding walls in shear ($H=0$) and add a purely oscillatory drift frequency which has four periods within the shear walls, $\od=(k_{y}\rho) K\cos[8\pi \l/L_{0}]$. We run linear simulations in GENE of a slab case $K=0$, a weaker curvature case $K=2\pi/L_{0}$, and a stronger curvature case $K=4\pi/L_{0}$. A scan is done over $k_{y}\rho$ and the mode with the peak growth rate is chosen. We plot in Figure \ref{fig:curv-onset}(a) the peak growth rate $\gamma$ in each case as a function of imposed temperature gradient. For small values of temperature gradient $(4\pi)^{-2} L_{0}/L_{T} < 0.5$, the absolute critical gradient for slab-like modes is observed. As the gradient increases a transition occurs to faster growing modes with higher $\kperp$ ($k_{y}\rho=0.4 \rightarrow 0.7)$, showing a ``knee'' \citep{Zocco2018} that is particularly visible for the stronger curvature case $K=4\pi/L_{0}$. In Fig. \ref{fig:curv-onset}(b) we plot the growth rate over the $k_{y}\rho$ spectrum for $K=4\pi/L_{0}$ at $4\pi^{-2}L_{0}/L_{T} > 1.59$ and $1.99$, showing the two different peaks of the spectrum and the higher $k_{y}\rho$ peak overtaking the low $k_{y}\rho$ peak. The knee feature thus captures the fact that the curvature-driven mode branch ($k_{y}\rho \simeq 0.7$)  has higher peak growth rates for these higher temperature gradients. For this case we extrapolate the stiffer behavior of peak $\gamma$ (between $4\pi^{-2}L_{0}/L_{T} = 1.99$ and $4\pi^{-2}L_{0}/L_{T} = 1.59$) back to an inferred critical gradient relating to the onset of the curvature-driven mode. The knee seems to nearly disappear when peak curvature is reduced by a factor of two (to $K=2\pi/L_{0}$), with an extrapolated curvature-induced critical gradient about a factor of two smaller. This is consistent with estimating the correlation length to be $\Lpar \propto R_{\text{eff,min}} = v_{T} k_{y}\rho/(\sqrt{2}\od_{\text{max}})$, the minimum effective radius of curvature. We see too that the knee emerges as $K$ is increased in the series of runs, for which $K\l_{\text{curv}}=0,\pi/4,$ and $\pi/2$, where $\l_{curv}$ is the width of one half-period of the cosine included in the curvature and is effectively the connection length. Since the knee starts to become visible in the weaker curvature case, $K\l_{curv}=\pi/4 \simeq 0.8$, we infer that $K\l_{curv} \gtrsim 1$ is required to observe the knee. This threshold condition could be related to the transition between slab-like and toroidal-like regimes of ETG turbulence discussed in \cite{Plunk2019}.

The transition between the fastest-growing-modes in the $K=4\pi/L_{0}$ case is made more explicit in Figure \ref{fig:curv-onset}(c) by plotting the mode structures of the peak growth rate modes before [$(4\pi)^{-2}L_{0}/L_{T}=1.59$, $k_{y}\rho=0.4$] and after [$(4\pi)^{-2}L_{0}/L_{T}=1.99$, $k_{y}\rho=0.7$] the knee. The mode structure before looks like a characteristic slab mode above marginality, with two nodes in the amplitude indicating it is probably a third-harmonic mode (the fundamental mode at marginality has no interior nodes as in Fig. \ref{fig:shearspikes}(a)). Once the transition occurs, however, the mode structure lines up neatly with the imposed curvature profile suggesting the mode is now sitting within bad curvature wells. The physics of an absolute critical gradient set by shear, and a second, curvature-induced critical gradient with a stiffer increase in growth rate can thus be modelled with bounding shear walls and an oscillatory curvature profile of fixed width. We note that our model geometry does not have consistency between the metrics and the drift curvature $\od$, as $\od$ should be related to $g^{yy}$ via $\bnabla y \cdot \bkappa$, but this feature allows us to study the effect of the terms separately. We now proceed to a case study demonstrating that the critical gradient can be strongly affected by manipulating the geometry of actual stellarator magnetic fields.

\section{Case study: controlling the critical gradient in a stellarator using field period number}\label{sec:helical}

Conventional wisdom holds that the ITG can be suppressed if the parallel extent of bad-curvature wells (connection length)  is reduced, as this would increase Landau damping of the modes that are localized to such regions. The inverse connection length is often approximated as $\kpar \sim 1/(qR)$ with R the major radius; this implies a characteristic damping frequency $\kpar v_{T}$ for the ITG, as frequently used since the late 1980s in tokamak research [e.g. \cite{Kim1991,Dominguez1989a}]. To test this idea in stellarators, one could manipulate a given equilibrium by increasing the toroidal field period number at constant aspect ratio so as to shorten all parallel length scales as the field periods are packed tighter into the torus \citep{Plunk2019}. We carry out a procedure like this below. As we shall soon see, however, what sets the ITG critical gradient in this case is global shear and not the local connection length as expected.

\subsection{Helical magnetic fields}

\begin{figure}
  \centering
  \scalebox{0.2}[0.2]{\includegraphics{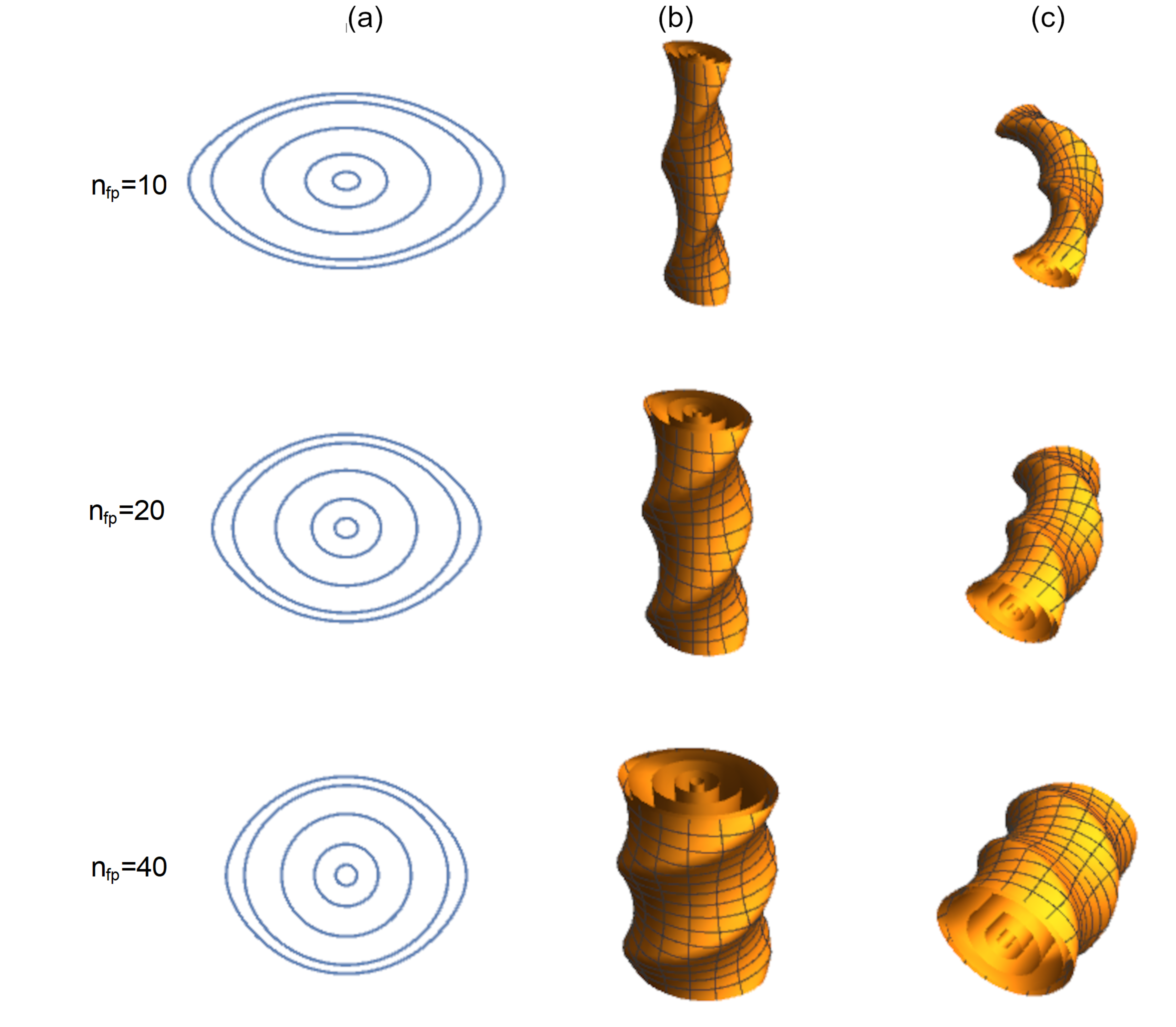}}
  \caption{Scan in helical field period number of the solutions to (\ref{eqn:fluxsurfaces}). Each row corresponds to a field period number $n_{\text{fp}}=n\zeta$ and each lettered column is a different plot. (a) Cross sections of each configuration for a variety of flux surfaces. (b) Extrusion of the helical cross-sections into cylindrical tubes with two field periods shown. (c) The same tubes as in (b) but now embedded into a toroidal shape.}
\label{fig:HelicalNFPScan}
\end{figure}

\begin{figure}
  \centering
  \scalebox{0.2}[0.2]{\includegraphics{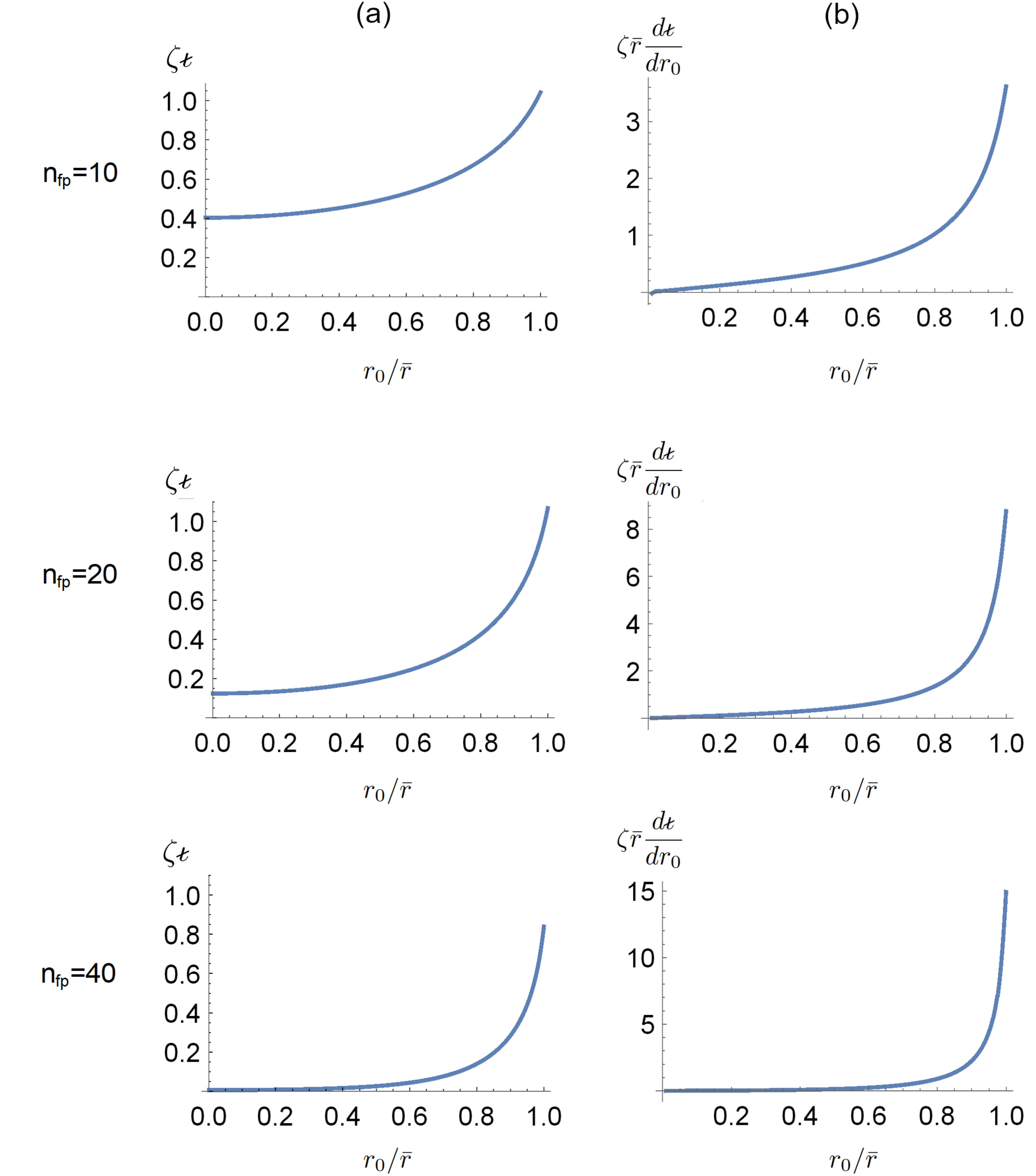}}
  \caption{Analytic results from the helical shapes in cylindrical geometry before embedding into a torus for each field period number $n_{\text{fp}}=n \zeta$. (a) Total rotational transform ( $\zeta \times \iotaslash$ of eqn. \ref{eqn:helicalomega}) as a function of $r_{0}/\bar{r}$. (b) $\bar{r}$ times the derivative of the quantity in (a) as a measure of global shear.}
\label{fig:HelicalNFPScan2}
\end{figure}

We begin with the helical solutions to Laplace's equation in cylindrical coordinates $(r_{\text{hel}},\phi,z)$ described in \cite{Morosov1966a} and employed in \cite{Bhattacharjee1983} to create ``straight'' stellarators. Note $\phi$ here is not to be confused with the electrostatic potential $\phi(\mathbf{x})$ introduced in section \ref{sec:definitions}. The solutions are eventually embedded into tori at large aspect ratio to produce the final configurations used to calculate the ITG linear critical gradient. The pre-embedded helical vacuum fields depend only on radius and the helical angle $\th=\phi - \zeta z$, where $\zeta=2\pi/L$ is the pitch of the field lines. $\theta$ is to be distinguished from the poloidal angle $\theta_{\text{pol}}$ defined in the introduction. $L$ is the total height of the cylindrical solution, with $n_{\text{fp}}= \zeta n$ the total number of helical field periods. The helical shapes are chosen in part because this continuous symmetry automatically implies quasisymmetry, which is only slightly broken during the embedding process. The solution of $\bnabla^{2}\Phi=0$ ($\mathbf{B}=\bnabla \Phi$) is written \citep{Morosov1966a}
\begin{equation}
    \Phi=B_{0}z+\frac{1}{\zeta}\sum_{n=1}^{\infty}b_{n} I_{n}(n \zeta r_{\text{hel}}) \sin(n\th),
\end{equation}
where $B_{0}$ is the constant magnetic field strength of $B_{z}$, $b_{n}$ are the perturbing magnetic field amplitudes that provide the rotation of the field lines, $I_{n}$ is a modified Bessel function and $n$ is the index of the perturbation harmonic. The vector potential $\mathbf{A}$ is found using $\pa \Phi/\pa r_{\text{hel}} = -\pa A_{\phi}/ \pa z $ and $(1/r_{\text{hel}})\pa \Phi/\pa \phi = \pa A_{r_{\text{hel}}}/ \pa z$, setting $A_{z}=0$. We choose a single helical perturbation $n=2$ and define the flux function as $\Psi=\zeta r_{\text{hel}} A_{\phi}$. The characteristic radius $r_{0}$ is defined through $\Psi_{0}=\zeta r^{2}_{0}B_{0}/2$. Setting $\Psi=\Psi_{0}$ then leads to
\begin{equation}\label{eqn:fluxsurfaces}
    \frac{\zeta^{2}}{2}(r_{\text{hel}}^{2}-r^{2}_{0})=\frac{b_{2}}{B_{0}}r_{\text{hel}} \frac{\mathrm{d} I_{2} (2 \zeta r_{\text{hel}})}{\mathrm{d} (2 \zeta r_{\text{hel}})}\cos(2\th),
\end{equation}
relating $r_{\text{hel}}$ to $\th$ for a flux surface $\Psi_{0}$. $n=2$ is chosen since it is the only helical perturbation that can sustain elliptical surfaces, and hence finite rotational transform, near the magnetic axis. With this choice (\ref{eqn:fluxsurfaces}) produces lens-shaped rotating flux surfaces, with ellipses at inner radii and cuspy edges, as the edge surfaces are bounded by a separatrix whose radius depends on $\zeta$ and $b_{2}/B_{0}$. Cross sections for these shapes are shown in Figures \ref{fig:HelicalNFPScan}(a), which are described in detail in Section \ref{sec:fieldperiod}. The rotational transform and global shear profiles of these shapes are presented in Fig. \ref{fig:HelicalNFPScan2}.

\subsection{Toroidal embedding}

\begin{figure}
  \centering
  \includegraphics[scale=0.52]{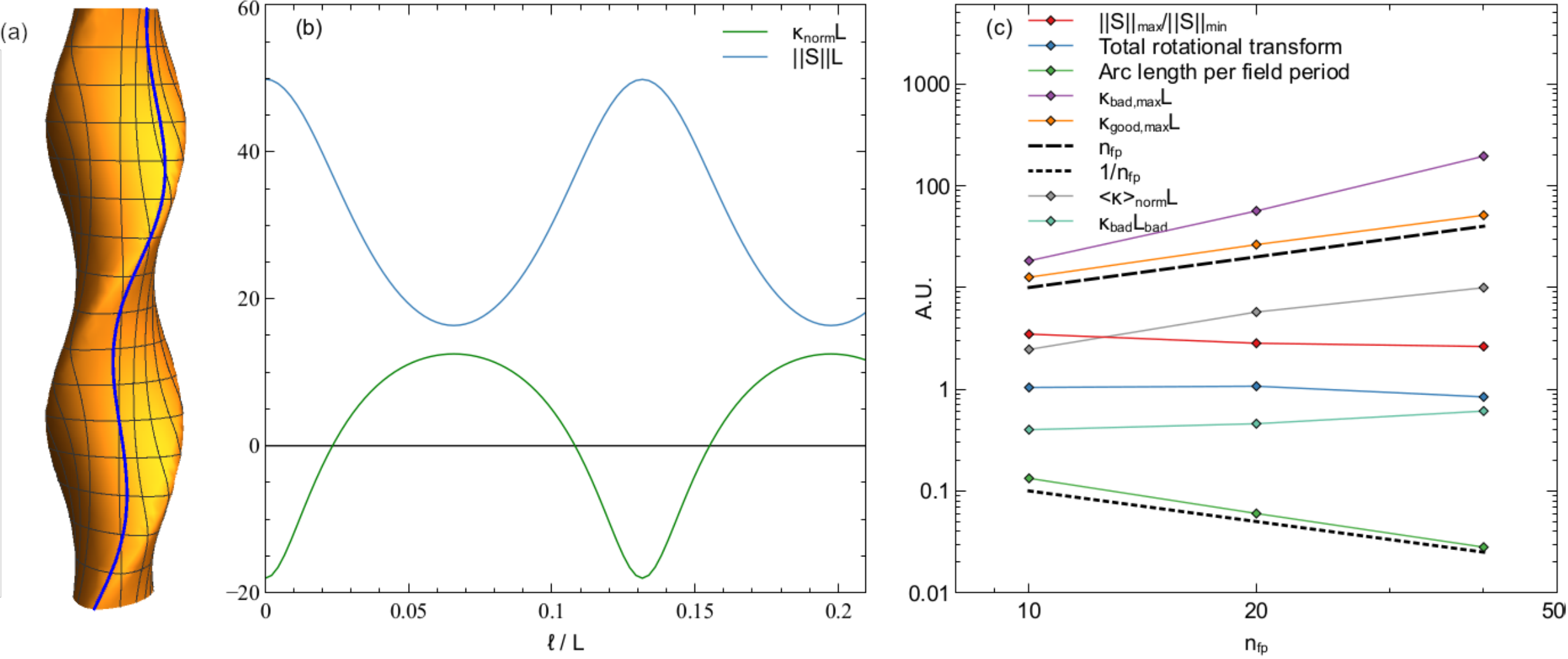}
  \caption{Field-line geometry calculations. (a) The surface $r_{0}/\bar{r}=0.9$ for $n_{\text{fp}}=10$ (orange) and the magnetic field line starting at $\phi=0,z=0$ (blue curve). (b) Normal curvature and shear amplification (equation \ref{eqn:locshear}) along the field line in (a). (c) Field line geometry outputs as a function of $n_{\text{fp}}$ with diamonds representing data points and curves connecting them on a log-log plot. The curves are: shear amplification factor $||S||_{\mathrm{max}}/||S||_{\mathrm{min}}$ (equation \ref{eqn:locshear}) (red), total surface rotational transform  $\zeta \times \iotaslash$ of eqn. \ref{eqn:helicalomega} (light blue), arc length per field period (green), maximum bad normal curvature (purple), maximum good normal curvature (orange), the line $y=n_{\text{fp}}$ (black dashed), the line $y=1/n_{\text{fp}}$ (black double dashed), average normal curvature (grey), and maximum bad curvature times the length of the bad curvature well (turquoise).}
\label{fig:HelicalGeomOutputs}
\end{figure}

To generate an equilibrium a fixed outer surface $r_{0}=\bar{r}$ that lies within the separatrix is chosen. This surface is then used to generate a toroidal field using a mapping, described as follows. The outer surface of the helical field, denoted by $r_{H}(\th,z)$ in the helical coordinate system, is first found by solving (\ref{eqn:fluxsurfaces}). The toroidal surface is then described by $x_{T}(\th,\phi_{T})=\hat{R}(\phi_{T}) R_T(\th,\phi_{T}) + \hat{z}_{T} Z_{T}(\th,\phi_{T})$, where $R_T(\th,\phi_{T})=R_{0}+r_{H}(\th,\phi_{T}/(2\pi))\cos(\th)$ and $Z_{T}(\th,\phi_{T})=r_{H}(\th,\phi_{T}/(2\pi))\sin(\th)$. The $z$-axis of the helical system is thereby mapped to a circle, with $\phi_{T}=2\pi z$. $\phi_{T}$ is a geometric toroidal angle not to be confused with the azimuthal angle $\phi$ in cylindrical coordinates. Note that the helical angle $\theta$ has become a geometric poloidal angle in these coordinates (it is a one-to-one mapping), and the $\hat{z}_{\text{T}}$ unit vector points vertically, aligned with the axis of the toroidal surface. We have also defined a major radius-like parameter $R_{0}=L/(2\pi)$. 

The toroidal surface is passed to VMEC \citep{Hirshman1983a} as input. The inverse aspect ratio of the torus is approximately $\eps=\bar{r}/R_{0}=1/10$, as $\bar{r}$ is effectively the minor radius of the outer surface of the VMEC solution. Since the inverse aspect ratio is small, most geometric quantities, such as rotational transform, are well approximated by the values of the helical solution. As such we approximate the scaling of important quantities such as arc length per field period and rotational transform in the final toroidal configurations using the analytic theory in cylindrical coordinates (Appendix \ref{app:helical}). In figure \ref{fig:HelicalGeomOutputs}(a) we show a field line on the surface $r_{0}/\bar{r}=0.9$ for $n_{\text{fp}}=10$. The field line alternates between regions of ``good'' (positive) and ``bad'' (negative) normal curvature $\bkappa \cdot \bnabla \Psi$ (Fig. \ref{fig:HelicalGeomOutputs}(b)).

\subsection{Scaling up field period}\label{sec:fieldperiod}

We now increase the field period number of the original helical equilibria by scaling up $\zeta$ but keeping $\bar{r}$ and $n=2$ fixed and reducing $b_{2}/B_{0}$ to maintain closed flux surfaces [$b_{2} \simeq 0.3/(\zeta dI_{2}(2\zeta r_{\text{hel}})/d(2\zeta r_{\text{hel}})|_{\bar{r}})$]. Reducing $b_{2}/B_{0}$ decreases poloidal derivatives and keeps edge rotational transform roughly constant, while increasing $\zeta$ at fixed minor radius scales up toroidal and radial derivatives as dictated by the vacuum solutions. The results of the field period scaling are shown in Figure \ref{fig:HelicalNFPScan} for $n_{\text{fp}}=n\zeta=10,20,$ and $40$, where each field period number is a row of the figure. Figures \ref{fig:HelicalNFPScan}(a) show the cross-sections of several surfaces in each configuration. At lower field period the surfaces are noticeably elliptical all the way inward to the axis, while the cross sections become circular as $\zeta$ is increased, except for cusp-like edges that persist near $\bar{r}$. Figures \ref{fig:HelicalNFPScan}(b) show two helical field periods of each configuration and the vertical compression of these field periods from larger field period number while Figures \ref{fig:HelicalNFPScan}(c) show the same two field periods after the toroidal embedding. We find in plots of the rotational transform (\ref{eqn:helicalomega}) and its derivative [Figs. \ref{fig:HelicalNFPScan}(c) and (d) respectively] that the strong radial derivatives at high field period suppress rotational transform in the core of the configurations and sharpen the cuspy edges, increasing global shear near the outermost flux surface. In addition, the regions of bad curvature get squeezed into small and smaller field line segments as field period is scaled up (Fig. \ref{fig:HelicalGeomOutputs}(c)).

As field period is increased, global shear at the edge becomes the dominant factor in setting the ITG mode correlation length and thus the critical gradient. Gyrokinetics in the sheared slab limit (cf \cite{Landreman2015}) tells us that, based on dimensionless ratios, the shearing length $L_{s}$ replaces $\lampar$ in the estimate (\ref{eqn:gabekadomtsev}) for the critical gradient such that $L_{s}/L_{T\text{crit}}=F$, where $F$ depends on dimensionless system parameters such as $\tau$. The shearing length, which comes from the secularly growing part of $g^{yy}$, can be estimated by fitting a parabola to the profile of $g^{yy}(\l)$ and setting the resulting coefficient of $\l^2$ to be $1/L^{2}_{s}$. This component increases with global shear and thus the critical gradient increases in proportion with toroidal field period number.

\subsection{GENE simulations near the outer flux surface}\label{sec:helicalGENE}

\begin{figure}
  \centering
  \includegraphics[scale=0.2]{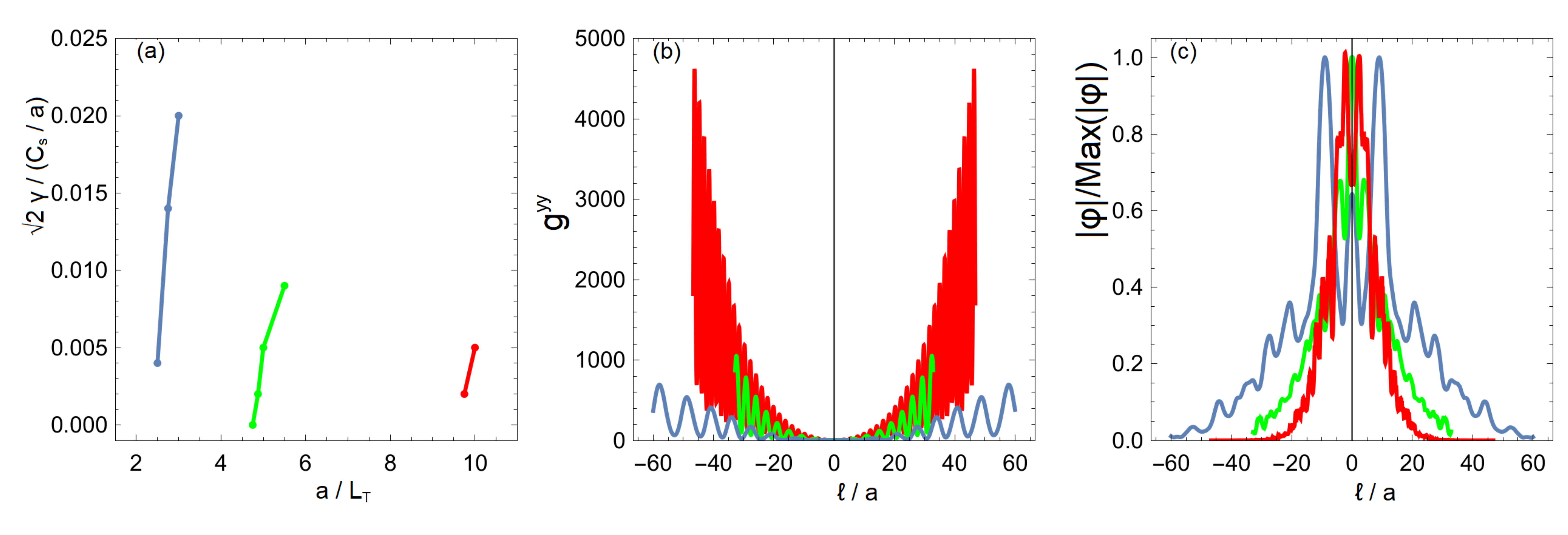}
  \caption{ITG linear simulations of toroidally embedded helical equilibria on the field line $\theta_{\text{pol}}=0$ on the surface with $99$ percent of the edge toroidal flux. The color coding is $n_{\text{fp}}=10$ (light blue), $20$ (green), and $40$ (red). (a) Scan in $a/L_{T}$ with $a$ the normalized average minor radius. Convergence to the linear critical gradient $\gamma \rightarrow 0$ for the last growing mode near marginality is found. (b) $g^{yy}$ plotted along the field line showing strong global shear as well as helical ripples. (c) $|\varphi|$ normalized to its maximum value along the field line showing reduction of the mode width as shear increases.}
\label{fig:HelicalGENECrit}
\end{figure}

The ITG linear critical gradients for the three equilibria $n_{\text{fp}}=10,20,40$ (Fig. \ref{fig:HelicalGENECrit}) are found by generating the solutions with the VMEC code \citep{Hirshman1983a}, passing them to the GIST package \citep{Xanthopoulos2009} and running the GENE code with the GIST output. In the GIST normalization, the temperature gradient and parallel length scales are now expressed relative to an effective minor radius $a \simeq \bar{r}$. As in section \ref{sec:shearspikes}, $T_{e}=T$, electrons are adiabatic, there is no density gradient and $m=0$ such that $k_{x}=0$. The standard twist-and-shift parallel boundary condition is used. We choose a flux tube in the outboard midplane $\theta_{\text{pol}}=0$ on the surface where (Toroidal flux / Edge Toroidal flux $=0.99$) for the three cases, with the toroidal angle defined to be zero, starting where the rotating cross section is level with the midplane. The ITG linear critical gradient is found by varying $a/L_{T}$ until $\gamma \rightarrow 0+$, where $a$ is the average minor radius and is a constant set in the GIST package. Here we do a scan in $k_{y}\rho$ to determine the global marginal mode. In figure \ref{fig:HelicalGENECrit}(a) we plot the results of $a/L_{T}$ scans in which the critical mode was found. We obtain a critical gradient $a/L_{T}\simeq 2.40$ for $n_{\text{fp}}=10$ which then scales almost exactly with field period $(a/L_{T}\sim 2,4,8)$. The profiles of $g^{yy}$ along the flux tube are plotted in Figure \ref{fig:HelicalGENECrit}(b), in which it becomes clear that the dominant length scale is set by the global shear parabola. The mode structures along the field line for the critical modes are plotted in Figure \ref{fig:HelicalGENECrit}(c). It can be inferred from these results that the lower bound on the critical gradient of the ITG mode, pertaining to broad-along-the-field-line, low-$k_{y}$ resonant modes, is set by global shear. Note that lower $k_{y}\rho$ modes than those predicted by the local slab theory (\ref{eqn:kadomtsev}) are to be expected, since $k_{y}\rho$ corresponds to the minimum value of $\kperp \rho$ and $\kperp \rho$ increases away from $\l=0$ when global shear is included. Thus small $k_{y}\rho$ is a signature of the sheared slab mode, as observed in GENE simulations of low-global-shear stellarators such as W7-X \citep{Zocco2018} and HSX \citep{Faber2018}. Since $n_{\text{fp}}=1/\epsilon=10$ retains rotational transform at the center of the device (fig. \ref{fig:HelicalNFPScan2} (a)), we speculate that setting $n_{\text{fp}}=1/\epsilon$ may help ITG mode stability, as such a configuration can still enjoy the benefits of reduced parallel length scales (compared to smaller field period numbers) while not increasing global shear too sharply throughout the volume.

\section{Direct calculation of the ITG linear critical gradient} \label{sec:critgrad-solve}
The integral equation (\ref{eqn:ballooning-disp}) includes the full physics of the linear electrostatic ITG critical gradient with adiabatic electrons. It also assumes $\mathrm{d}|\mathbf{B}|/\mathrm{d}\l=0$ such that particle trapping is neglected and all ions can be treated as passing. The integrals contained can be challenging to evaluate as they contain a singularity and are oscillatory. Eq. (\ref{eqn:ballooning-disp}) has been solved for the case of a circular tokamak geometry \citep{Romanelli1989a} assuming finite $\gamma$ and a density gradient. We start by setting $\gamma = \Im [\omega]=0$ and as mentioned in section \ref{sec:gyrotunnel} curvature now enters into the problem by modifying the phase factor $\lambda$ which we define as
\begin{equation}\label{eqn:mbarav}
\lambda(\ell^{\prime}, \ell)\equiv \sgn(\ell-\ell^{'})M(\ell^{\prime}, \ell) =\frac{|\ell-\ell^{'}|}{\xpar \vth}(\o-\left<\odt \right>),
\end{equation}
where eq. (\ref{eqn:mbar}) was used and $\left< \cdot \right>$ denotes an average over the parallel length between $\ell$ and $\ell^{'}$, i.e. 
\begin{equation}
    \odtav(\l,\lp)=\frac{1}{|\l - \lp|} \int_{\lp}^{\l}\mathrm{d}\ell^{\prime\prime} \,\odt(\ell^{\prime\prime}).
\end{equation}
If $\odtav = \odav(
\xpar^2+\xperp^2/2)$ is positive (``average bad curvature''), cancellation with $\o$ can happen, leading to a decrease of $\lambda$ and thus an increase in the effective correlation length $v_{T}/\lambda$ through the drift resonance. This is the primary drive mechanism for the toroidal branch of the ITG and leads to the onset of resonant modes driven by curvature explored in section \ref{sec:curv-onset}. Since the case $\odav < 0$ (average ``good curvature'') eliminates the resonance (we use the sign convention $\osT,\o>0$), it is expected to provide uniform stabilization of the mode by increasing the magnitude of the phase factor and effectively decreasing the correlation length $v_{T}/\lambda$. However, even $\odtav>0$ can lead to damping of the mode as mentioned in section \ref{sec:curv-onset} by pulling the mode out of the drift resonance once $\odtav > \o$. Thus in the cases of $\lambda$ large and of either sign, curvature can result in a damping effect from decorrelation similar to that of the parallel slab resonance (section \ref{sec:kadomtsev}).

Note that the effect of curvature on the drive of the mode may not be reflected in the mode width. For instance, the average curvature can be roughly uniform and have no outward effect on the mode structure, but it may affect the drive of the mode, especially for an extended slab-like mode as tends to be seen near marginality. We thus caution that the mode width and the distance between local features in the geometry (nominal connection length) are both potential ``red herrings'' that generally do not reveal the correlation length $\Lpar$. $\Lpar$ is a ``hidden'' length captured by the integral equation (\ref{eqn:ballooning-disp}) that can be less than the mode width. With some basic understanding of the physics of curvature in hand we now outline the steps to solving (\ref{eqn:ballooning-disp}).

\subsection{$\xperp$ Integration}

Some analytical progress can be made in (\ref{eqn:ballooning-disp}) by performing the $\xperp$ integral first \citep{Romanelli1989a} using Weber's formula,
\begin{align}\label{eqn:weber}
    \int_{0}^{\infty}\mathrm{d}\xperp \xperp J_{0}\left(\sqrt{2b(\ell)} \xperp \right)J_{0}\left(\sqrt{2b(\ell^{'})} \xperp \right)\exp{(-p \xperp^{2})} \nonumber \\  =\frac{\exp(-(\bl+\blp)/(2p))}{2p}I_{0} \left(\frac{\sqrt{\bl \blp}}{p} \right) \equiv \hat{\Gamma_{0}}(\bl,\blp,p)
\end{align}
where the grad-B drift has been absorbed into the factor $p=1+i \, \odb/(2\xpar \vth)$, with $\odb \equiv |\ell-\ell^{'}|\left<\od \right>$. The singularity at $\xpar=0$ in $p$ causes no obvious issues since Weber's formula is valid for $\mathrm{Re}[p]>0$, which is guaranteed by the $1$ coming from the Maxwellian in \xperp. The term $\propto \xperp^3$ from the $\osT$ factor can also be integrated analytically by writing it as a $p$ derivative of (\ref{eqn:weber}) yielding what we call $\goh$ (Appendix \ref{app:weber}).

The $\hat{\Gamma}$ functions contain interactions between curvature and shear. Curvature can provide additional oscillatory behavior to the integral through the imaginary term in $p$ and can suppress the effect of shear in the exponent and $I_{0}$ if $1/p$ is very small, although this tends to be the limit where phase mixing from $\ob$ already suppresses the integral for $\odav < \o$. $\gzh$ behaves similarly to a Dirac delta function $\delta((\bl - \blp)/p)$, causing large suppression if $\bl$ and $\blp$ are very large and different but giving a finite contribution to the integral if $\bl \simeq \blp$ or $1/p<\bl - \blp$. These observations follow from the asymptotic form of $I_{0}$ at large argument.

\subsection{$\xpar$ Integration}

Equation \ref{eqn:ballooning-disp} can now be written

\begin{align}\label{eqn:ballooning-disp-xperp}
\varphi(\ell) = \frac{-i}{\vth\sqrt{\pi}}\int_{-\infty}^{\infty}d\ell^{\prime}\varphi(\ell^{\prime})\int_{0}^{\infty} \frac{d\xpar}{\xpar} (\o \gzh - \osT((\xpar^2-3/2)\gzh+\goh) \nonumber \\ 
\times   \exp\left[-\xpar^2 + i \left(\frac{\ob}{\xpar \vth}-\frac{\odb \xpar}{\vth} \right)\right],
\end{align}
where $\ob \equiv |\ell-\ell^{'}|\o$. The $\xpar$ integration can also be done analytically in the case $p=1$ ($\od=0$) when written as Meijer-G functions times $\gzh$ or $\goh$, although the logarithmic singularity at $\lp=\l$ and $\xpar=0$ must still be avoided in the $\lp$ integration.

Returning to the general case including curvature, we note that, for $\l \ne \lp$, the $\xpar^{-1}$ singularity is shielded by the oscillatory behavior in the $\ob$ exponent and also negated by the factor $1/p \rightarrow 0$ for $\xpar \rightarrow 0$ and $\odav \ne 0$ in (\ref{eqn:ballooning-disp-xperp}). In numerical solutions of the $\xpar$ integration (we know of no analytical expression for the integral including \od) the $\xpar$ integrand at $\l = \lp$ is replaced by the centered average of its values at $\lp= \l \pm  \xi \, \delta \l$,  where $\delta \l$ is the spacing of the parallel coordinate grid and $\xi = 1/8$ for the results presented in section \ref{sec:tanhwall-solutions}.

\subsection{$\lp$ Integration}

We now discretize $\varphi(\l)$ in order to evaluate the integral in (\ref{eqn:ballooning-disp-xperp}) by assuming that $\varphi$ is piecewise constant over $N$ sub-domains, with $\l$ set to be at the center point of each sub-domain. This allows us to pull the constant $\varphi$ out and integrate the remaining kernel numerically across each sub-domain, yielding an $N \times N$ matrix $\mathbf{G}$ such that (\ref{eqn:ballooning-disp-xperp}) becomes $\varphi_{i}=G_{ij}\varphi_{j}$. The numerical integration is done over an $\lp$ grid with $N^{'}$ points. The $\l$ grid overlaps with the $\lp$ grid but we use a finer resolution in $\lp$ such that $N^{'}\ge 2N+1$. This reduces the total number of integral evaluations by a factor $N^{'}/N$ compared to a scheme where $N=N^{'}$ but can resolve basic mode structures, such as a mode that peaks near $\l=0$ and decays for large enough $|\l|$, as could be modelled with $N \geq 3$. The scheme still captures some of the effect of fine-scale features in the geometry through the $\lp$ integration.

Equation (\ref{eqn:ballooning-disp-xperp}) is finally written as
\begin{equation}\label{eqn:matrix-ballooning-disp}
    \mathrm{Det}[\mathbf{G}-\mathds{1}]=0
\end{equation}
where $\mathds{1}$ is the identity matrix of size $N$. Note that condition (\ref{eqn:matrix-ballooning-disp}) is satisfied only if there is an eigenvector with zero eigenvalue, since the determinant is the product of the eigenvalues.

\subsection{Eigenvalue problem} \label{sec:tanhwall-solutions}

\begin{figure}
  \centering
  \includegraphics[scale=0.45]{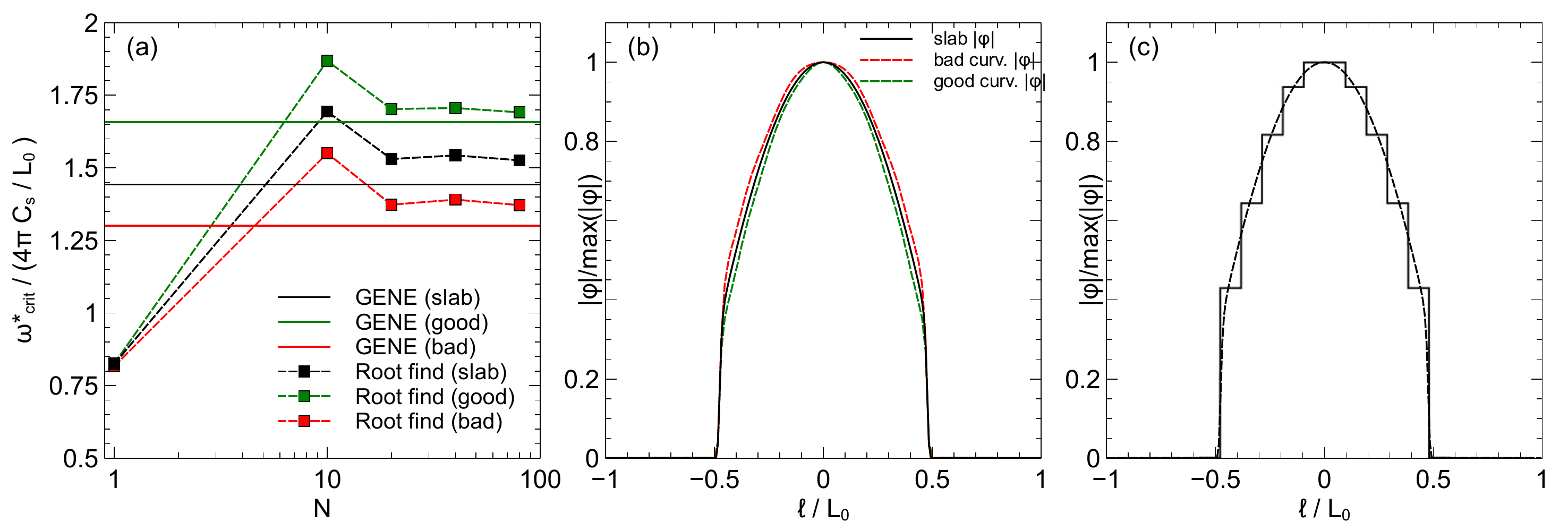}
  \caption{Validating solutions of (\ref{eqn:ballooning-disp-xperp}) against GENE simulations. (a) Scan in $N$, the number of points in $\l$, used for the solution of (\ref{eqn:ballooning-disp-xperp}) showing convergence of the eigenvalue $\osT_\text{crit}$ (squares connected by dashed lines) for the shear geometry of section \ref{sec:shearspikes} with varying constant curvature, where black is the slab case with no curvature, green is with ``good'' curvature $\Reff=-L_{0}/2$ and red is the ``bad'' curvature case $\Reff=L_{0}/2$. Values taken from GENE are plotted as solid horizontal lines for validation of the root find. (b) Mode structures from GENE for the three geometries in (a) with the same color scheme and dashed lines for cases with curvature. (c) Step function representation of $|\vp|$ from the solution of (\ref{eqn:ballooning-disp-xperp}) with $N=20$ overlaid on the GENE solution of $|\vp|$ (dashed) for the slab case with no curvature.}
\label{fig:root-convergence}
\end{figure}

Equation (\ref{eqn:matrix-ballooning-disp}) is traditionally considered an eigenvalue problem, with the complex quantity $\o$ as the eigenvalue. Here we reinterpret it, treating the pair of real frequencies $(\o,\osT)$ as the eigenvalue. It is not clear {\it a priori} whether the integral equation should be well-behaved at $\gamma=0$ (since it was derived assuming $\gamma>0$ as in \cite{Connor1980}). We find, however, that the determinant smoothly approaches zero and a root can be found to eight digits of precision, proving the concept for finding the critical $\osT$ (and hence critical gradient) as an eigenvalue of the dispersion relation. We use a numerical Newton solver to obtain roots to this equation for three test cases: the sheared slab geometry plotted in \ref{fig:shearspikes}(a) (equation \ref{eqn:shearspikes}), the same case but with constant ``bad'' curvature $\od=\sqrt{2}k_{y}\rho v_{T}/(2L_{0})$, and the case with constant ``good'' curvature $\od=-\sqrt{2}k_{y} \rho v_{T}/(2L_{0})$, where the effective radius of curvature in the latter two cases is $\Reff=k_{y} \rho v_{T}/(\sqrt{2}\od)=L_{0}/2$. The mode has $k_{y}\rho=\kperp\rho=0.8$, i.e. no sweep in $k_{y}$ space is performed to find the true critical mode, but the change in the onset of the single, marginal ITG mode with $k_{y}\rho=0.8$ is tracked. A root is found for the values $N=1,10,20,40,80$, ($1<N<10$ does not always converge on a sensible answer), keeping a fixed $N^{'}=161$, for the three cases. 

In Figure \ref{fig:root-convergence}(a) we plot the eigenvalues $\osT$ against the critical gradients obtained from linear GENE simulations with the same shear profile and the appropriate constant curvature added. For large $N$ the curves show agreement with the GENE critical gradients to within $6\%$. The convergence is non-uniform (fluctuating about the presumed fully converged answer) since the numerical method uses three grids to evaluate the $\lp$ integral, only one of which is refined in this series by increasing $N$. There is uncertainty in the critical gradients read off from the GENE simulations owing to finite precision of the GENE solution and the requirement of finding a mode above marginality ($\gamma>0$). As such we consider convergence to within $6\%$ deviation from the GENE critical gradient to be acceptable.

As expected in the case of a modified slab mode with $\o > \odav$, curvature shifts the critical gradient for $k_{y0}=0.8$ up (good curvature) or down (bad curvature) by roughly $10\%$, which is comparable to the ratio $\odav/\o_{\mathrm{slab}}\simeq 1/5$. The plots of $|\vp(\l)|$ from GENE confirm that the slab solution mode structure is barely affected by the curvature [Fig. \ref{fig:root-convergence}(b)]. For the solutions with $N=1$ in Figure \ref{fig:root-convergence}(a), very little difference between good, bad, and zero curvature cases is seen and the critical gradients from the root finds are about half the converged answers. A large deviation for $N=1$ is not surprising since the requirement that $\varphi=0$ at the boundaries in $\l$ is not even satisfied by a constant $\varphi$. In Figure \ref{fig:root-convergence}(c) we show the approximate eigenfunction for $N=20$ overlaid on the GENE solution for the zero curvature case to visualize the discretization scheme. The choice $N=20$ captures the rapid decay of the eigenfunction near $|\l|/L_{0}=1/2$ particularly well.

A convergence check for the $N=10$ bad curvature case was also performed with finite positive gamma to ensure that no discontinuities between $\gamma=0$ and $\gamma > 0$ were present, e.g. delta function terms $\propto \delta(\gamma)$ that appear in quasilinear theory of the ITG mode \citep{Helander2018}. A root was found after setting a fixed $\gamma/|\o|\simeq 0.01$, where $\o$ is that of the $\gamma=0$ solution. A corresponding increase of one percent in both $\o$ and $\osT$ relative to the case of $\gamma=0$ occurred, suggesting the solutions are smooth near $\gamma=0$. It may be that an averaging effect from integrating over many points in $\lp$ helps regularize the integral for $\gamma=0$.

\section{Discussion}\label{sec:conclusion}

In this work, the gyrokinetic ITG mode linear critical gradient has been investigated using analytic theory and numerical models in flux tube geometries. We found that traditional connection length estimates often do not adequately capture the physics of the onset of the linear electrostatic ITG mode, and that instead the correlation length $\Lpar$, which emerges in the solution of the integral equation (\ref{eqn:ballooning-disp}), is the appropriate reference length that sets the critical gradient through the damping frequency $v_{T}/\Lpar$. The correlation length can be determined by local shear, global shear, curvature, or some combination of all three of these geometric properties. It has long been assumed that strong local shear features inherent to stellarators will stabilize ITG turbulence. However, in section \ref{sec:shearspikes} we discovered using numerical experiments that large shear amplification factors of the perpendicular wavenumber $\kperp(\l)$ with broad extent along the field line are needed to significantly increase the ITG mode linear critical gradient. Using this model as a basis, we then showed the onset of curvature-driven resonant ITG modes whose correlation length is set by the peak magnitude of drift curvature in section \ref{sec:curv-onset}. Subsequent experiments in section \ref{sec:helical} revealed that the absolute critical gradient for the ITG can be predictably increased through global shear via field period number using helical stellarator shapes. The optimal field period number may be close to the toroidal aspect ratio of the configuration, since for larger field periods the benefits to the ITG mode are localized to small radial layers. Motivated by this evidence, we turned to direct calculation of the critical gradient in section \ref{sec:tanhwall-solutions} through solving equation (\ref{eqn:ballooning-disp}), in which we observed that average magnetic drift over the mode extent is likely the most important factor in setting the critical gradient aside from global shear. 

We note that several approximations were made in solving equation (\ref{eqn:ballooning-disp}) that constrain the validity of the solution, such as neglecting particle trapping and electromagnetic effects on the ITG mode associated with finite plasma $\beta$ \citep{Zocco2015d}. However, trapped particles may ultimately contribute not very strongly to the linear behavior of the ITG \citep{Proll2013a}, especially if this particle population is small. This may occur in optimized stellarators since reducing trapped particle fractions is one way to lower neoclassical transport \citep{Dinklage2018}.

The method we have devised to calculate the linear critical gradient has several potential advantages over running a traditional GK linear solver. First, simulations become increasingly difficult approaching the critical gradient, due to the large timescales required for convergence whereas there is no time variable for our eigenvalue problem. The numerical grid for GK solvers generally also includes two velocity coordinates, whereas our approach integrates over these variables, leaving a single variable, the field line following coordinate. While we can only speculate for now, we suspect an optimized solution of equation \ref{eqn:ballooning-disp} using tabulated integrals and a decent initial guess for $(\o,\osT)$ should take about a second per value of $k_{x}$ or $k_{y}$ on a single processor for $N=10$ points used in the discretization of the eigenmode structure. For reference, the parallelized GENE simulation close to the critical gradient depicted in fig. \ref{fig:shearspikes}a took about 20 seconds for the marginal $k_{y}\rho$, and this was the result of a sweep downward from higher guesses of $a/L_{T}$.

The simple numerical model developed in section  \ref{sec:GENEmodels} and used to validate the solver in section \ref{sec:tanhwall-solutions} doesn't use the true geometry of stellarators but does contain field-line varying metric coefficients. As such, while a stellarator would have more complicated variation along the field line, the evaluation of the geometry inputs would be the same as in our simple model. The possibility of a rapid calculation of the critical gradient as a general figure of merit for ITG turbulence is an appealing prospect for stellarator optimization. Work toward this goal is now underway.

\section*{Acknowledgements}
G.T.R.C. would like to thank M.J. Pueschel, T. G\"orler, and A. Ba\~non Navarro for essential help with running the GENE code, as well as A. Zocco for helpful discussions and comments on the manuscript. We also thank the anonymous reviewers for their help in improving the clarity and content of the manuscript. The GENE simulations were carried out on the draco cluster located at IPP Garching in Germany. This work was supported by a grant from the Simons Foundation (No. 560651, P. H.).

\appendix

\section{Ballooning theory}\label{ballooning-appx}

Eqn.~\ref{gk-eqn} is differential at first order in $\ell$ and can be recast in integral form by taking the right-hand-side as a source and integrating directly, as shown by \cite{Connor1980}.  This is done for each sign $\sigma = \mbox{sign}(\vpar)$ separately, because we will need to use $\sigma$-dependent boundary conditions.  Rewriting Eqn.~\ref{gk-eqn} using $\vpar = \sigma|\vpar|$ we have

\begin{equation}
\frac{\partial g}{\partial \ell} - i\frac{\sigma}{|\vpar|}(\o - \odt)g = -i\frac{\sigma}{|\vpar|}(\o - \ost)J_0\varphi f_0\label{gk-eqn-2}
\end{equation}

\noindent We multiply this equation by the quantity $\exp(-i\sigma M(\ell_0, \ell))$, where

\begin{equation}
M(\ell_0, \ell) = \int_{\ell_0}^{\ell} \frac{\o - \odt(\ell^{\prime})}{|\vpar(\ell^{\prime})|} d\ell^{\prime},
\end{equation}

\noindent and integrate to arrive at the expression

\begin{equation}
g(\ell) = g_0 \exp(i\sigma M(\ell_0, \ell)) - i\sigma(\o - \ost)f_0\int_{\ell_0}^{\ell}\frac{J_0^{\prime}}{|\vpar^{\prime}|}\varphi(\ell^{\prime})\exp(i\sigma M(\ell^{\prime}, \ell))d\ell^{\prime},\label{g-soln-0}
\end{equation}

\noindent where $\vpar^{\prime} = \vpar(\ell^{\prime})$ and $J_0^{\prime} = J_0(k_{\perp}(\ell^{\prime})v_{\perp}(\ell^{\prime})/\Omega)$.

To fix the constant of integration $g_0$, consider $\Im[\o] = \gamma > 0$ and $\ell_0 = -\sigma \infty$, then note that exponential divergence of the term $g_0 \exp(-i \sigma M)$ will occur unless $g_0 = 0$.  This is equivalent to applying the boundary conditions

\begin{eqnarray}
&g(\vpar > 0, \ell = -\infty) = 0,\label{bound-cond-a}\\ 
&g(\vpar < 0, \ell = \infty) = 0.\label{bound-cond-b}
\end{eqnarray}
Eqn.~\ref{g-soln-0} then becomes

\begin{align}
g(\ell) = - i\sigma(\o - \ost)f_0\int_{-\sigma \infty}^{\ell}\frac{J_0^{\prime}}{|\vpar^{\prime}|}\varphi(\ell^{\prime})\exp(i\sigma M(\ell^{\prime}, \ell))d\ell^{\prime}.\label{g-soln-1}
\end{align}

\noindent Substituting this into Eqn. (\ref{qn-eqn}) yields Eqn. (\ref{eqn:ballooning-disp}). This equation is also derived in \cite{Romanelli1989a} following \cite{Connor1980}.

\section{Weber Integral}\label{app:weber}

The term including the factors $\osT \xperp^{3}$ in equation (\ref{eqn:ballooning-disp}) is 

\begin{align}
    \goh(\bl,\blp,p)=\int_{0}^{\infty}\mathrm{d}\xperp \xperp^{3} J_{0}\left(\sqrt{2b(\ell)} \xperp \right)J_{0}\left(\sqrt{2b(\ell^{'})} \xperp \right)\exp{(-p \xperp^{2})} \nonumber \\  =-\frac{\mathrm{d}\gzh}{\mathrm{d}p}= \frac{\exp(-(\bl+\blp)/(2p))}{p^{3}}\left[-\left(\frac{\bl+\blp}{2}-p\right)I_{0}\left(\frac{\sqrt{\bl \blp}}{p} \nonumber \right) \right. \\ 
    \left. + \sqrt{\bl\blp}I_{1}\left(\frac{\sqrt{\bl \blp}}{p} \right) \right] 
\end{align}
where $\gzh$ is defined in equation (\ref{eqn:weber}).

\section{Helical equilibria calculations}\label{app:helical}
\subsection{Rotational transform}

The average rotational transform on a helical surface can be computed from the equation of a field line, $r \mathrm{d}\phi/\mathrm{d}r=B_{\phi}/B_{r}$. We quote the result in \cite{Morosov1966a} for $n=2$,

\begin{align}\label{eqn:helicalomega}
    \hat{\iotaslash}=\frac{\delta \phi}{\delta \phi + \pi/2}  \\ 
    \delta \phi = \int_{r_{\mathrm{min}}}^{r_{\mathrm{max}}} \mathrm{d}r_{\text{hel}} \frac{ (r_{\text{hel}}^{2}-r_{0}^2)(I_{2}/(dI_{2}(2\zeta r_{\text{hel}})/d(2\zeta r_{\text{hel}}))) } {\zeta r_{\text{hel}}^{2} \sqrt{ \left(\frac{2 r b_{2} dI_{2}(2\zeta r_{\text{hel}})/d(2\zeta r_{\text{hel}})}{\zeta B_{0}} \right)^{2}-(r_{\text{hel}}^{2}-r^{2}_{0})^{2}   } },
\end{align}

where $\hat{\iotaslash}$ is the rotational transform per two helical field periods (called $\o$ in \cite{Morosov1966a}), $r_{\mathrm{max}}$ is the maximal surface radius (where $\th=0$), $r_{\mathrm{min}}$ is the minimal radius ($\th=\pi/4$), and $\delta \phi$ is the total azimuthal displacement of a field line as $\th$ varies by $-\pi/2$. Note that oscillatory behavior is superimposed on the secular increase of $\phi$ per field period and averaged out by the integration.

\subsection{Coordinate-free representation of the geometry}
The analytic solutions for the helical shapes (\ref{eqn:fluxsurfaces}) can be used to find geometric quantities. For instance, we can calculate a proxy measure for local shear from $\mathbf{S}=\bnabla \hat{b}$, where $\bhat$ is the unit magnetic field vector (not to be confused with $b=k^{2}_{\perp}$). Noting that $\hat{b} \cdot \bnabla \al = \hat{b} \cdot \bnabla \Psi = 0$, we find
\begin{equation}\label{eqn:shearamp}
 \bhat \cdot \bnabla k^{2}_{\perp}\equiv k^{2}_{\perp 0}\bhat \cdot \bnabla f(\l)=-\mathbf{k}_{\perp}\mathbf{k}_{\perp}:\mathbf{S}
\end{equation}
where $\mathbf{k}_{\perp 0}=\mathbf{k}_{\perp }(\l=0)$ and $f(\l)=k^{2}_{\perp}(\l)/k^{2}_{\perp 0}$ is the shear amplification factor. Thus we can find $f(\l)$ by extracting the perpendicular components of $\mathbf{S}$, defining
\begin{equation}
-\mathbf{k}_{\perp}\mathbf{k}_{\perp}:\mathbf{S} \equiv k^{2}_{\perp 0}\left(\cos^{2}(\vt)S_{xx} +\sin^{2}(\vt)S_{yy}+ \cos(\vt)\sin(\vt)(S_{xy}+S_{yx})\right).
\end{equation}
In this notation a proxy for the upper bound on amplification by local shear (which would operate in tandem with the secular increase from global shear) is 
\begin{equation}\label{eqn:locshear}
||\mathbf{S}||=\sqrt{S^{2}_{xx}+S^{2}_{xy}+S^{2}_{yx}+S^{2}_{yy}}=\sqrt{(\bnabla_{\perp}\bhat):(\bnabla_{\perp}\bhat)}=\sqrt{(\bhat \times \mathbf{S}):(\bhat \times \mathbf{S})}
\end{equation}
where the final two definitions don't require magnetic coordinates to be defined. In the case of the helical surfaces (\ref{eqn:fluxsurfaces}), only the equation for a field line in cylindrical coordinates must be solved and derivatives taken. $\mathbf{S}$ also yields the variation of normal curvature ($\bkappa \cdot \bnabla \Psi$) along the field line since $\bkappa=\hat{b} \cdot \mathbf{S}$, the sign of which can tell whether curvature will be favorable or unfavorable for the toroidal branch of the ITG mode. $\mathbf{S}$ also has the well-known property that $\mathbf{S}\cdot \bhat=0$.

\bibliographystyle{jpp}
\bibliography{library.bib}

\end{document}